\documentclass[%
 aip,
 amsmath,amssymb,
 reprint,%
]{revtex4-1}

\usepackage{graphicx}
\usepackage{dcolumn}
\usepackage{bm}

\usepackage[utf8]{inputenc}
\usepackage[T1]{fontenc}
\usepackage{mathptmx}
\usepackage{etoolbox}
\usepackage{xcolor}

\makeatletter
\def\@email#1#2{%
 \endgroup
 \patchcmd{\titleblock@produce}
  {\frontmatter@RRAPformat}
  {\frontmatter@RRAPformat{\produce@RRAP{*#1\href{mailto:#2}{#2}}}\frontmatter@RRAPformat}
  {}{}
}%
\makeatother
\begin{document}

\preprint{AIP/123-QED}

\title[Dynamics of Meniscus-Bound Particle Clusters in Extensional Flow]{Dynamics of Meniscus-Bound Particle Clusters in Extensional Flow}
\author{Sagar Chaudhary}
\affiliation{Department of Mechanical Science and Engineering, University of Illinois at Urbana-Champaign, Urbana, IL 61801}
\affiliation{Beckman Institute for Advanced Science and Technology, University of Illinois at Urbana-Champaign, Urbana, IL 61801}
\author{Sachin S. Velankar}%
 \affiliation{Department of Chemical and Petroleum Engineering, University of Pittsburgh, Pittsburgh, PA 15260}
\author{Charles M. Schroeder}
\email{cms@illinois.edu}
\affiliation{Beckman Institute for Advanced Science and Technology, University of Illinois at Urbana-Champaign, Urbana, IL 61801}
\affiliation{Department of Chemical and Biomolecular Engineering, University of Illinois at Urbana-Champaign, Urbana, IL 61801}
\affiliation{Department of Materials Science and Engineering, University of Illinois at Urbana-Champaign, Urbana, IL 61801}
\affiliation{Center for Biophysics and Quantitative Biology, University of Illinois at Urbana-Champaign, Urbana, IL 61801}
\date{\today}

\begin{abstract}
Capillary suspensions are three-phase mixtures containing a solid particulate phase, a continuous liquid phase, and a second immiscible liquid forming capillary bridges between particles. Capillary suspensions are encountered in a wide array of applications including 3D printing, porous materials, and food formulations, but despite recent progress, the micromechanics of particle clusters in flow is not fully understood. In this work, we study the dynamics of meniscus-bound particle clusters in planar extensional flow using a Stokes trap, which is an automated flow control technique that allows for precise manipulation of freely suspended particles or particle clusters in flow. Focusing on the case of a two-particle doublet, we use a combination of experiments and analytical modeling to understand how particle clusters rearrange, deform, and ultimately break up in extensional flow. The time required for cluster breakup is quantified as a function of capillary number $Ca$ and meniscus volume $V$. Importantly, a critical capillary number $Ca_{crit}$ for cluster breakup is determined using a combination of experiments and modeling. Cluster relaxation experiments are also performed by deforming particle clusters in flow, followed by flow cessation prior to breakup and observing cluster relaxation dynamics under zero-flow conditions. In all cases, experiments are complemented by an analytical model that accounts for capillary forces, lubrication forces, hydrodynamic drag forces, and hydrodynamic interactions acting on the particles. Results from the analytical models are found to be in good agreement with experiments. Overall, this work provides a new quantitative understanding of the deformation dynamics of capillary clusters in extensional flow. 
\end{abstract}

\maketitle


\section{\label{sec:intro}Introduction}

Capillary state suspensions are formed by adding a second immiscible liquid to a particle suspension, resulting in the formation of particle clusters bound by liquid capillary bridges \cite{koos2014capillary, koos2011capillary}. Capillary suspensions are a remarkably simple class of materials with tunable properties and are encountered in a wide range of applications such as low fat food formulations, 3D printing, porous ceramics, and conductive pastes for printed electronics \cite{schneider2017suppressing,amoabeng2018bulk,koos2012tuning, hoffmann2014using, roh20173d,dittmann2013ceramic,studart2006processing, schneider2016highly}. In addition, the transition of a mixture from a fluid-like two-phase suspension to a gel-like capillary suspension \cite{koos2011capillary} can be used to overcome issues related to particle sedimentation \cite{yang2019magnetorheological} and to enhance yield stress \cite{koos2014capillary, van1975rheology}. Despite their importance in industrial applications, mixing operations for preparing capillary suspensions remain largely empirical \cite{hauf2018structure, bossler2017influence}. From this view, we a lack a complete understanding of the rheology and fluid mechanics required to control and manipulate the dynamic microstructure of capillary suspensions in flow. 

The wetting of solid particles by a second liquid leads to clustering, resulting in the formation of liquid-bound particle clusters that serve as the building blocks of capillary suspensions. Early experimental work focused on measuring the capillary force on particles in the presence of a liquid bridge. MacFarlane and Tabor calculated the force of adhesion due to surface tension between a flat plate and a bead covered in a thin film of liquid for a stationary system \cite{mcfarlane1950adhesion}. In 1965, Mason and Clark designed a method to measure the forces exerted on two polyethylene hemispheres by liquid bridges of a di-$n$-butyl phthalate/liquid paraffin mixture suspended in water \cite{mason1965liquid}. Mason and Clark used relatively large particles (radius $R$ = 1.5 cm) to facilitate facile and accurate experimental measurements because interparticle forces are expected to increase proportionately with particle size. Over the last few decades, improvements in experimental techniques have allowed for more accurate measurements of smaller magnitude capillary forces. Willett \emph{et al}. measured capillary forces using a sensitive microbalance interfaced with a computer, enabling measurement of a wide range of particle radii and bridge volumes \cite{willett2000capillary}. 

The discussion thus far has been restricted to static forces wherein the particle motion is sufficiently slow to be regarded as quasi-static. Shifting focus to forces between moving particles, Pitois \emph{et al}. investigated the effects of meniscus viscosity on capillary and viscous forces for small bridge volumes between two moving spheres \cite{pitois2000liquid}. Displacement-controlled experiments were performed where meniscus-bound spheres were mechanically separated until rupture, and the forces were measured as a function of separation distance and sphere velocity. Analytical expressions for capillary and viscous forces were developed and compared with the experiments. Recently, Bozkurt and coworkers studied the effects of wettability and meniscus volume on the capillary force between two equally sized moving glass beads using a similar displacement controlled approach. Capillary force was measured for various particle radii and separation velocities, and the experimental data were compared with analytical models in literature \cite{pitois2000liquid, bozkurt2017capillary, likos2004unsaturated}. Theoretical work on liquid bridges has also been performed to estimate capillary forces. Early work dates back to 1926, when Fisher approximated the shape of the bridge as a toroid \cite{fisher1926capillary}. Following this work, Derjaguin used an approximation where the meridian radius of the toroidal shaped bridge is much smaller than the radius of the neck, yielding an analytical expression for the capillary force \cite{derjaguin1934untersuchungen, israelachvili2022surface}. The Derjaguin equation was subsequently used by several researchers in the field for capillary suspensions \cite{rabinovich2005capillary, butt2009normal}. However, the toroidal and Derjaguin approaches for calculating capillary forces are only valid for small bridge volumes and small interparticle distances \cite{lian1993theoretical}, resulting in significant errors at large separations. Motivated by these limitations, several researchers have developed more accurate expressions for the capillary force in liquid bridges \cite{rabinovich2005capillary, maugis1987adherence, pitois2000liquid}.

Broadly, prior work on capillary state suspensions has largely focused on either quasi-static particle clusters or displacement-controlled motion of particles, where the relative motion between particles is prescribed by a mechanical system. However, this experimental scenario is not representative of mixing flows encountered by freely suspended particles in capillary suspensions where the relative motion of particles is dominated by hydrodynamic forces exerted by the surrounding fluid. For these reasons, we lack a complete understanding of the fundamental rheology and fluid mechanics in capillary state suspensions which will help to inform the design of mixing operations in bulk systems. Key questions include: what is the characteristic timescale required for a liquid-bound cluster to rupture in flow? What is the effect of the bridge volume or particle wettability on cluster dynamics and breakup? And, what is the role of interparticle hydrodynamic interactions in cluster breakup in flow? Despite recent progress, the dynamics of freely suspended liquid-bound particle clusters in flow is not fully understood.

Unlike particle clusters, flow-induced breakup has been extensively studied for liquid droplets. Research on the breakup dynamics of immiscible liquid droplets traces back to 1934, when G.I. Taylor studied the deformation and bursting of a drop in a second immiscible liquid under various controlled flow conditions, generating extensional flow using a manually controlled “four roller” apparatus and shear flow using a “parallel band” apparatus \cite{taylor1934formation}. Taylor developed criteria for drop deformation as a function of dimensionless flow strength (capillary number, $Ca$) \cite{taylor1934formation}. Several decades later, Grace investigated droplet breakup in extensional and shear flow, resulting in the classic Grace curve which plots critical capillary number $Ca_{crit}$ for drop breakup as a function of the viscosity ratio of the suspension \cite{grace1982dispersion}. Bentley and Leal used a computer-controlled four-roll mill to study elongation and breakup of small liquid droplets in two-dimensional linear flows \cite{bentley1986experimental}. Experiments using the four-roll mill were performed for various flow types, and drop shapes were compared with predictions from several burst theories \cite{bentley1986experimental}. Numerical studies on shape relaxation and capillary breakup of initially extended drops under no-flow conditions have also been conducted for various initial drop shapes \cite{stone1989relaxation}. Despite the seemingly vast amount of prior literature on liquid drop dynamics, far less insight is available for the micromechanics of meniscus-bound particle clusters. In fact, many of the same physical questions arise for particle clusters bound by liquid bridges; for instance, what is the relaxation behavior of an initially extended cluster? What is the critical capillary number $Ca_{crit}$ required for cluster breakup? And what is the effect of the viscosity of the bridge fluid relative to the bulk fluid? Although these questions have been resolved for liquid droplets decades ago, they have not yet been addressed for meniscus-bound particle clusters in external flows. Recent advances in automated flow control \cite{shenoy2016stokes, kumar2019orientation, shenoy2019flow, kumar2020automation, tu20233d} provide an ideal approach to study the dynamics of freely suspended particle clusters in precisely controlled flows.

In this work, we investigate the dynamics of meniscus-bound particle doublets (two-particle clusters) in planar extensional flow using a Stokes trap \cite{shenoy2016stokes}, which allows for the manipulation of particles in precisely defined flows. Meniscus-bound particle doublets (clusters with two particles) are the basic unit comprising capillary suspensions. Experiments are complemented by analytical models that incorporate the effects of capillary forces, lubrication forces, hydrodynamic drag forces, and hydrodynamic interactions acting on the particles. Our results show that particle cluster doublets undergo reorientation and eventual breakup in extensional flow above a critical capillary number $Ca_{crit}$. The analytical model is used to predict $Ca_{crit}$ for cluster breakup, as well as to understand the effects of capillary number $Ca$ and meniscus volume $V$ on breakup time. Particle cluster relaxation is further studied under zero-flow conditions revealing insights regarding capillary and viscous modes of relaxation. In all cases, experimental results are in good agreement with the analytical model. Taken together, our results present a fundamental understanding of particle cluster dynamics in external flows that could be useful for designing and implementing mixing processes for capillary suspensions. 

\begin{figure*}[t]
\includegraphics[width=0.75\textwidth]{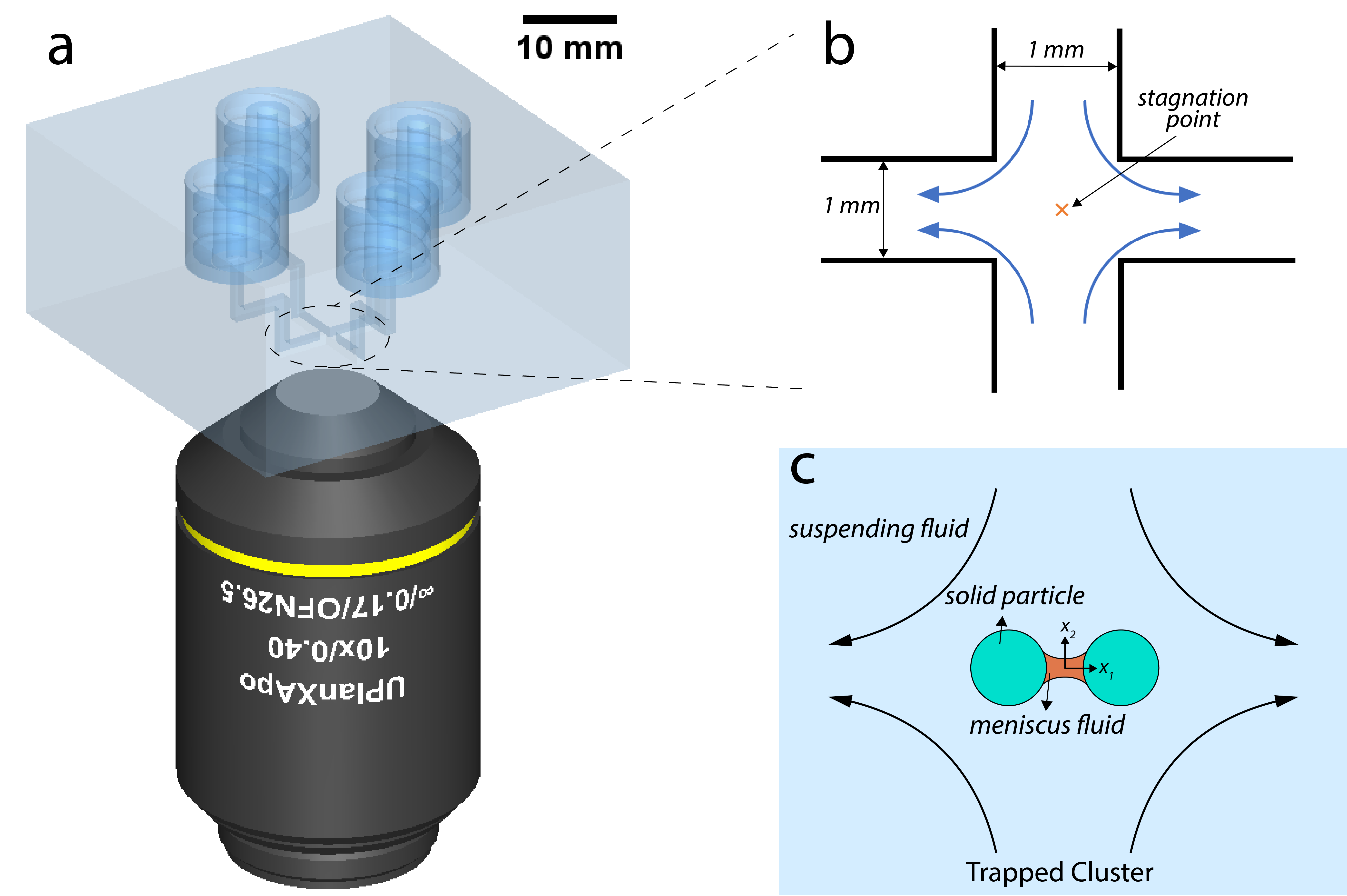}
\caption{\label{fig:devicemat} Schematic of microfluidic device used to study particle cluster dynamics in extensional flow. (a) Schematic of the 4-channel microfluidic device fabricated using SLA 3D printing along with a 10$\times$ magnification air-immersion objective lens. The height and width of the channels is 1 mm. The four channels are connected to fluid reservoirs pressurized by regulators and controlled using a LabVIEW program. The cross-slot geometry generates a planar extensional flow, and the open surface of the device is sealed with a glass coverslip for optical microscopy. (b) Schematic showing a zoomed-in view of the 4-channel cross-slot in the microfluidic device. The curved arrows depict the extensional flow, and the stagnation point is labeled near the center of the cross-slot. Channel width is 1 mm. (c) Schematic showing a two-particle meniscus-bound cluster trapped in extensional flow at the stagnation point. The suspending fluid, solid particles, and the meniscus fluid are labeled. The radius of the particles is 50 $\mu$m.}
\end{figure*}

\section{\label{sec:methods}Materials and methods}
\subsection*{Microfluidic Device and Stokes Trap}
An automated flow technique known as the Stokes trap \cite{shenoy2016stokes, kumar2019orientation, shenoy2019flow, tu20233d} is used to study the dynamics of meniscus-bound particle clusters in flow. We use a four-channel microfluidic device with a cross-slot geometry at the center to generate planar extensional flow (Figure \ref{fig:devicemat}). Using the Stokes trap, meniscus-bound particle clusters are confined near the stagnation point in extensional flow for long times or accumulated fluid strain. The microfluidic device is fabricated using a stereolithography (SLA) 3D printer (Form 3, Formlabs) utilizing a photopolymer clear resin (V4, Formlabs). Devices are printed with 1 mm-wide by 1 mm-tall channel dimensions using a previously reported procedure by Tu \emph{et al}. \cite{tu20233d}. The bottom surface of the microdevice is sealed with a glass coverslip to allow for optical microscopy, as previously reported \cite{tu20233d}. The Stokes trap is used in conjunction with an inverted microscope (Olympus IX71) coupled with a CMOS camera (Grasshopper 3, FLIR), with the entire setup placed on a vibration-damped optical table (Thorlabs). The microfluidic device is mounted on the microscope stage, and a 10$\times$ magnification air-immersion objective lens (0.4 NA, numerical aperture) is used to image into the cross-slot area (Figure \ref{fig:devicemat}(a)). All four inlet/outlet channels of the device are connected via polytetrafluoroethylene (PTFE) tubing to fluid reservoirs, three of which contain only the bulk suspending fluid, and the remaining channel contains a low concentration of particle clusters in the bulk suspending fluid. The reservoirs are pressurized using high-precision pressure regulators (Elveflow OB1MK3+) to establish and control a pressure driven flow in the device. A model predictive control (MPC) algorithm \cite{shenoy2016stokes} is used together with a custom LabVIEW interface to control the flow field and manipulate and trap particle clusters near the stagnation point of planar extensional flow in the center of the cross-slot channel.

Numerical simulations of the flow geometry are performed using COMSOL Multiphysics to model the flow of a Newtonian fluid (viscosity 10 Pa$\cdot$s) in a 3D model of the device. Channel dimensions (width and height) are set to 1 mm to match the device dimensions in experiments. Two of the four channels are designated as inlets, and the remaining two channels are assigned as outlets. A no-slip boundary condition is specified for solid surfaces in the entire geometry, and the velocity profile is solved using the Stokes equations, yielding a zero velocity stagnation point at the center of the cross-slot.

\begin{figure*}[t]
\includegraphics[width=0.9\textwidth]{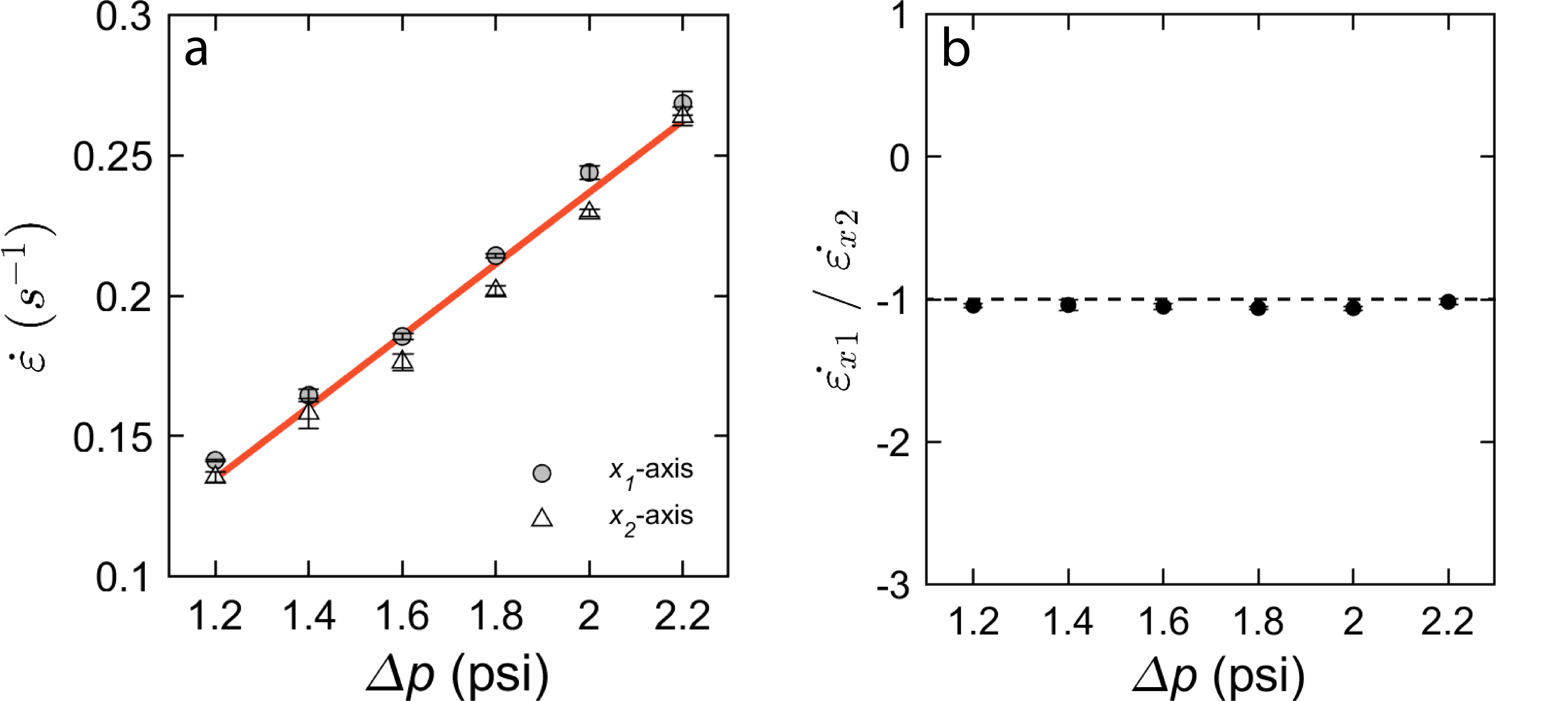}
\caption{\label{fig:flowchar} Flow characterization in the cross-slot of the microfluidic device. (a) Strain rate, $\dot{\epsilon}$ at the central plane of the microfluidic device is determined as a function of the input pressure, $\Delta p$ in the channels for both $x_1$ and $x_2$ directions. Error bars show standard deviation from multiple ($\ge$3) experiments. Polystyrene (PS) particles (10 $\mu$m in diameter) are added to silicone oil ($\eta_c$=10 Pa$\cdot$s) for particle tracking experiments. (b) Ratio of the strain rates in $x_1$ and $x_2$ directions is determined to be $\approx$-1, thus verifying the existence of planar extensional flow in the cross-slot ($\dot{\epsilon}_{x_1}=-\dot{\epsilon}_{x_2}$).}
\end{figure*}

\subsection*{Materials}
Silicone oil (density $\rho_c$ = 0.971 g/mL, Sigma-Aldrich) is used as the suspending fluid, polyisobutene (PIB-24, density $\rho_d$ = 0.89 g/mL, Soltex) is used as the meniscus fluid, and polyethylene (PE) spheres (100 $\mu$m diameter, Cospheric) are used as the solid particulate phase. Materials selection is based on a density matching condition to avoid significant sedimentation of particle clusters in the no-flow direction ($x_3$-direction). The meniscus fluid is immiscible with the bulk fluid and wets the particles with an average contact angle $\theta=$ 55$^o$. A shear rheometer (DHR-2, TA Instruments) was used to measure the viscosity of the suspending and meniscus fluids ($\eta_c$ = 10 Pa$\cdot$s for silicone oil and $\eta_d$ = 19 Pa$\cdot$s for PIB-24). The selection of high viscosity fluids facilitates access to large capillary numbers $Ca$, as discussed below. The surface tension between the two fluids was measured as $\sigma$ = 3 mN/m. PE particles were imaged using environmental scanning electron microscopy (ESEM) to measure their diameter; the average particle diameter was 100 $\pm$ 5 $\mu$m and ranged between 90 $\mu$m to 106 $\mu$m. Meniscus-bound clusters are prepared using bulk mixing. Here, a small volume of the meniscus fluid PIB-24 (approximately 100 $\mu$L) is mixed with 1 mL of silicone oil to obtain a milky white dispersion of fine droplets. Subsequently, 100 mg of PE particles are added to the dispersion and gently mixed, allowing for collisions between particles and PIB-24 drops, thus forming capillary bridges between particles. The mixture is further diluted (at a ratio of 1:100) to obtain the final mixture used for imaging and experiments. Figure \ref{fig:devicemat}(c) shows a schematic of two-particle meniscus-bound cluster in planar extensional flow, trapped near the stagnation point at the center of the cross-slot in the microfluidic device. During imaging experiments, we routinely observe particle clusters with variable number of particles $N=2,3,4,$ and higher order clusters, however in this work, we focus only on studying particle doublets (two-particle clusters), using the Stokes trap setup. 

\subsection*{Particle Tracking Velocimentry}
Particle tracking velocimetry (PTV) is used to characterize the flow field in the microfluidic device. Fluorescent polystyrene (PS) particles (10 $\mu$m diameter, ThermoFisher Scientific) are used for PTV experiments. The aqueous sample is first centrifuged to obtain the dense phase of PS particles, which are then resuspended in silicone oil (10 Pa$\cdot$s viscosity). The suspension is then sonicated (Branson 2510R-MT) for 45-50 minutes to obtain a well mixed particle sample for experiments. The PS particles are introduced into the device, and videos ($\approx$5 min in duration) of particles convecting in planar extensional flow are captured in the vicinity of the cross-slot as a function of strain rate $\dot{\epsilon}$. Video files (supplementary material, Movie 1) are used as an input for image analysis where particle velocities and trajectories are determined using a custom MATLAB program, as previously described \cite{tu20233d}. Particle velocities in the $x_1$ and $x_2$-directions are determined as a function of position, and the data are fitted to the flow field equations for planar extensional flow using the velocity gradient tensor as a fitting parameter:
\begin{equation} \label{eq:flowvel}
\begin{bmatrix}
v_{x_1}\\
v_{x_2}
\end{bmatrix} = \begin{bmatrix}
a & b\\
c & d
\end{bmatrix}
\begin{bmatrix}
x_1\\
x_2
\end{bmatrix}
\end{equation}
The matrix entries $a$ and $d$ yield the strain rates in $x_1$ and $x_2$-directions ($\dot{\epsilon}_{x_1}$ and $\dot{\epsilon}_{x_2}$) respectively and the entries $b$ and $c$ are found to be approximately zero. Using this approach, $\dot{\epsilon}_{x_1}$ and $\dot{\epsilon}_{x_2}$ are determined as a function of the input pressure in the microfluidic system (Figure \ref{fig:flowchar}(a)). Three PTV experiments are performed at each of the input pressure values, and average strain rate $\dot{\epsilon}$ and standard deviation values are reported. Figure 2(b) shows a plot of the ratio of the strain rate in the $x_1$-direction to the strain rate in $x_2$-direction as a function of inlet pressure, which further validates the existence of planar extensional flow (i.e., $\dot{\epsilon}_{x_1}=-\dot{\epsilon}_{x_2}$).

\begin{figure}[t]
\includegraphics[width=0.45\textwidth]{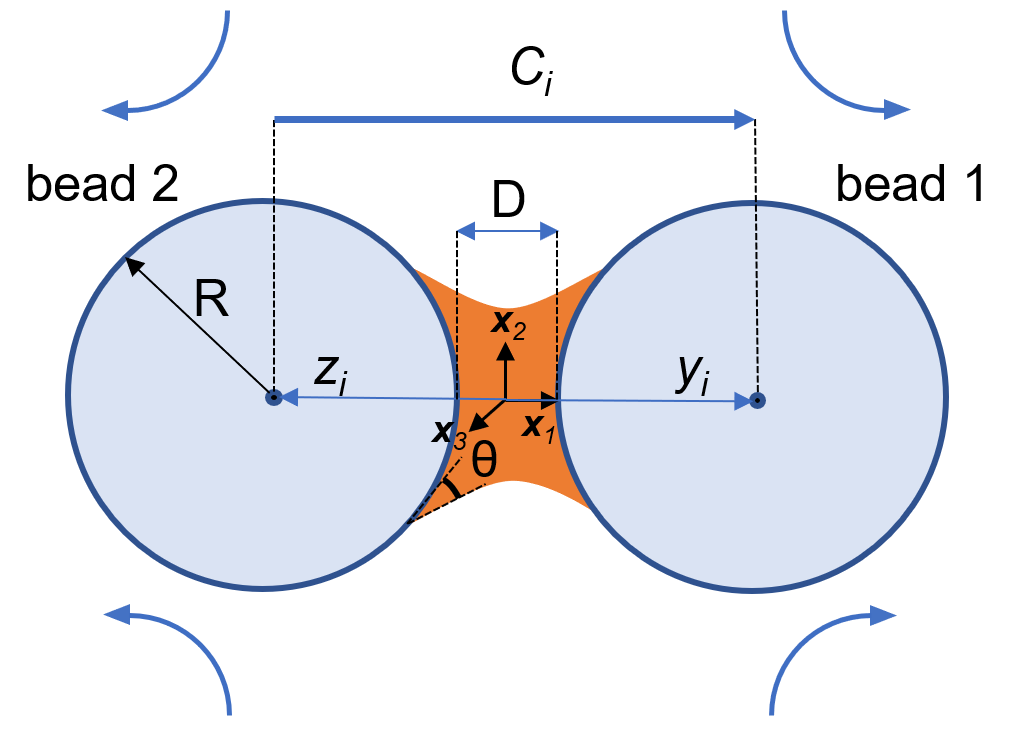}
\caption{\label{fig:clusterschem} Schematic of a two-particle cluster bound by a liquid bridge showing relevant parameters and the coordinate system. $R$ is the particle radius, $C_i$ is the vector pointing from the center of particle 2 to the center of particle 1, $D$ is the smallest separation between the particle surfaces, $y_i$ and $z_i$ are the vectors pointing from the origin to the centers of particles 1 and 2, respectively, and $\theta$ is the contact angle.}
\end{figure}

\section{\label{sec:model}Analytical Model}
\subsection*{Cluster Breakup}
We developed an analytical model for particle cluster breakup that accounts for the forces acting on the particles as well as interparticle hydrodynamic interactions (HI). Figure \ref{fig:clusterschem} shows a schematic of a two-particle liquid-bound cluster in planar extensional flow showing the relevant variables and coordinate system. The cluster breakup phenomenon in extensional flow depends on the interplay between the hydrodynamic drag forces, capillary forces, and lubrication forces between particles. All of our experiments are performed in the limit of low Reynolds numbers (particle Reynolds number, $Re_p = \rho_c R U / \eta_c \approx 10^{-7}$; and device Reynolds number, $Re_d = \rho_c H U / \eta_c \approx 10^{-6}$, where $R$ is the particle radius, $H$ is the channel height of the device, and $U$ is a characteristic fluid velocity), hence inertia is neglected. In the following, we use index notation such that subscripts refer to the vector or tensorial nature of a quantity. A force balance on particle $j$ in a two-particle cluster yields:
\begin{equation} \label{eq:forcebal}
F^{cap,j}_i + F^{vis,j}_i = 0 \quad \quad \quad \quad j=1\; \text{or} \; 2
\end{equation}
where $F^{cap,j}_i$ is the capillary force on bead $j$ and $F^{vis,j}_i$ is the net viscous force on bead $j$. Depending on the interparticle separation, the mathematical model for cluster breakup is described by two regimes. In the first regime, the interparticle separation $D$ is relatively small ({\em i.e.}, $D/R \lessapprox 1$), such that particles experience lubrication forces, a hydrodynamic drag force from the flowing fluid, and an attractive capillary force from the liquid bridge. In the second regime, the interparticle separation $D$ is relatively large ($D/R \gtrapprox 1$) such that particles experience a hydrodynamic drag force from the fluid and a capillary force from the bridge until breakup. Consequently, we present two models, referred to below as the near-field and far-field models for cluster breakup in extensional flow. Motivated by prior work \cite{pitois2000liquid, bozkurt2017capillary, lopez2023low}, we consider the lubrication approximation as valid for interparticle separations up to $D/R=0.5$. Thus, for cluster breakup, the near-field model is considered for $D/R<0.5$,  whereas for $D/R\geq0.5$, the far-field model is used.  

\subsubsection*{Near-field Model}
The liquid bridge connecting the two particles applies an attractive capillary force that acts to pull them together \cite{lian2016capillary}. An expression for the capillary force acting on a particle due to a small liquid bridge with a fixed volume was derived by Maugis \cite{maugis1987adherence} and reproduced by Pitois \emph{et al}. \cite{pitois2000liquid}. Using this result, the capillary force acting on bead 1 is:
\begin{equation} \label{eq:2}
F^{cap,1}_i = -2\pi R\sigma (\cos \theta) \left[1 - \frac{1}{\sqrt{1 + \frac{2V}{\pi RD^2}}} \right] \hat{C}_i
\end{equation}
where $\sigma$ is the interfacial tension, $\theta$ is the contact angle, $V$ is the liquid bridge volume, $D$ is the smallest separation between the two particle surfaces, and $\hat{C}_i=C_i / |C_i|$ is the unit vector along the direction of separation between bead centers, as shown in the schematic in Figure \ref{fig:clusterschem}. From Figure \ref{fig:clusterschem}, also note that $y_i - z_i=C_i$.

Particles experience a net viscous force arising from the hydrodynamic drag force due to the straining motion of the suspending fluid and due to lubrication forces between particles for small interparticle separations ($D/R < 0.5$) \cite{pitois2000liquid, bozkurt2017capillary, lopez2023low}. The net viscous force on bead $j$ is given by:
\begin{equation} 
F^{vis,j}_i = F^{drag,j}_i + F^{lub,j}_i
\end{equation}
where $F^{drag,j}_i$ is the hydrodynamic drag force exerted on bead $j$ due to flow and $F^{lub,j}_i$ is the lubrication force acting on bead $j$. In a transient deformation experiment, the initial state of a two-particle cluster corresponds to a small spacing between adjacent particle surfaces separated by a thin liquid film. Lubrication forces are hence expected to play a key role in determining viscous forces between particles and controlling particle dynamics. Using the Reynolds equation, Lopez \emph{et al}. \cite{lopez2023low} calculated the pressure profile over the entire domain (meniscus and bulk suspending fluid), which was then integrated to obtain the total viscous force acting on a particle by the two fluids. Using this result, the lubrication force acting on particle 1 in the presence of a liquid bridge is:
\begin{multline} \label{eq:3}
F^{lub,1}_i= - \left[ \frac{6\pi\eta_{c} R^2 }{\sqrt{\frac{V}{\pi R} + D^2}}+ \frac{6\pi R^2 }{D}\left(\sqrt{\frac{V}{\pi R}+ D^2}-D\right) \right.\\
\left. \times \frac{\left(\eta_{d} \sqrt{\frac{V}{\pi R}+ D^2} + D(\eta_{c}-\eta_{d})\right)}{\frac{V}{\pi R}+ D^2}\right] \frac{dD}{dt} \hat{C}_i
\end{multline}
where $\eta_d$ is the meniscus fluid viscosity. Note that equation (\ref{eq:3}) includes the viscous drag force due to the center-of-mass velocity of the bead \cite{lopez2023low}. In this work, we consider cluster breakup in a planar extensional flow, which is described by a velocity field $v_i = E_{ij} x_j$, where $E_{ij} = \dot{\epsilon}(\delta_{i1}\delta_{j1} - \delta_{i2}\delta_{j2})$ is the rate-of-strain tensor and $x_j$ is a vector pointing from the origin. For a two-particle cluster aligned along the extensional axis, the hydrodynamic drag force on particle 1 due to the flowing fluid is approximated by Stokes' law:
\begin{equation} \label{eq:1}
F^{drag,1}_i \doteq 6\pi R\eta_{c}\dot{\epsilon} y_i
\end{equation}
where $R$ is the particle radius, $\dot{\epsilon}$ is the strain rate, $y_i$ is the vector from the origin to the center-of-mass of bead 1, and $\eta_{c}$ is the viscosity of the bulk suspending fluid. Equation (\ref{eq:1}) provides a reasonable approximation to the hydrodynamic drag force exerted on the particle in the doublet. As an aside, we validated this approximation by solving the analytical cluster breakup model while treating the numerical coefficient in Stokes' law as a fitting parameter. This analysis resulted in a good agreement between the experiments and the model and yielded a value for the numerical coefficient as $\approx$5.8, which is close to the value in Stokes drag for a sphere, thus justifying the approximation of hydrodynamic drag force using Eq. (\ref{eq:1}).

Particles also experience hydrodynamic interactions (HI) due to flow field disturbances generated by nearby particles in the cluster. Our experimental results show that the cluster breakup phenomenon in extensional flow is sensitive to small perturbations in fluid velocity, and thus we anticipate that HI plays a key role in particle cluster breakup dynamics. Here, the disturbance velocities $v_i^D$ are calculated, and the hydrodynamic drag force is modified to account for these interactions. In the following, we consider the disturbance flow generated by particle 2 acting on particle 1 (Figure \ref{fig:clusterschem}). The first term in a multipole expansion of the disturbance velocity is the Stokeslet \cite{graham2018microhydrodynamics}, which arises due to the capillary force exerted on the particles. The disturbance velocity experienced at bead 1 due to the presence of bead 2 is given by the Stokeslet:
\begin{equation} \label{eq:4}
v_{i\:(Stokeslet)}^{D,1}= G_{ij}F_{j}
\end{equation}
where the superscript $D$ denotes disturbance. Here, $G_{ij}$ is the Oseen-Burgers tensor \cite{graham2018microhydrodynamics} and $F_{j}$ is the force exerted on the fluid by the adjacent particle (taken here as bead 2) due to the capillary force. The Oseen-Burgers tensor $G_{ij}$ is written as:
\begin{equation} \label{eq:5}
G_{ij} = \frac{1}{8\pi\eta_{c} C} \left(\delta_{ij} + \frac{C_i C_j}{C^2} \right)
\end{equation}
where $C$ is the center-to-center distance between particles, $C_i$ is the vector pointing from the center of particle 2 to the center of particle 1, and $\delta_{ij}$ is the Kronecker delta. The two-particle cluster is taken to be aligned along the extensional flow axis ($x_1$-axis), such that the vector $C_i$ is oriented in the positive $x_1$ direction and has magnitude $|C_i|=|C_1|=C$. In order to calculate the disturbance velocity experienced by bead 1 due to bead 2, $F_{j}$ is taken as the force exerted on the fluid by bead 2, which is equal to the negative of the net viscous force ($-F^{vis,j}_i$) exerted on bead 2, or the positive of the capillary force on bead 2 (see Eq. (\ref{eq:forcebal})), such that:
\begin{equation}\label{eq:6}
F_{j}^{cap, 2} = F^{cap}\hat{C}_j  
\end{equation}
where $F^{cap}$ is the scalar magnitude of the capillary force defined in equation (\ref{eq:2}) and $\hat{C}_j$ is the unit vector in the $x_1$ direction. The expression for the Stokeslet disturbance velocity experienced by particle 1 becomes:
\begin{equation} \label{eq:7}
v_{i\:(Stokeslet)}^{D,1} = \frac{F_{cap}}{4\pi\eta_{c} C} \hat{C}_i
\end{equation}
The Stokeslet contribution to the disturbance velocity assumes a point force in flow. The potential dipole term in the multipole expansion would account for the finite size of a sphere translating under an imposed force. However, as a simplifying assumption, this term is not included in the disturbance velocity because the particle translational velocity is relatively small, and the potential dipole term is therefore negligible in magnitude compared to other dominant terms in the disturbance velocity such as the stresslet (vide infra).

In addition to the Stokeslet, we also account for the disturbance flow due to the straining motion of the suspending fluid by introducing the stresslet \cite{graham2018microhydrodynamics}. In addition, the interparticle separation is typically smaller than the particle size ($D/R < 1$) at the start of the breakup process, and we therefore need to account for the finite size of the particles by introducing the potential quadrupole \cite{graham2018microhydrodynamics}. The stresslet and potential quadrupole terms both arise due to the imposed straining flow field. Consider the case of a finite sized sphere held fixed in a general linear flow in the creeping flow regime. The total fluid velocity $v_{i}^{total}$ for this flow can be written as \cite{graham2018microhydrodynamics}:
\begin{equation} \label{eq:8}
v_{i}^{total} = v_{i}^{\infty} + v_{i}^{D} = v_{i}^{\infty} + \frac{p' x_i}{2\eta_c} + v_{i}^{H}
\end{equation}
where $v_{i}^{\infty} = E_{ij} x_{j}$ is the undisturbed fluid velocity, $v_{i}^{D}$ is the disturbance velocity, $p'$ is the dynamic pressure, and $v_{i}^{H}$ is a harmonic function. The rate-of-strain tensor for planar extensional flow $E_{ij}$ is given in matrix form as:
\begin{equation} \label{eq:9}
E_{ij} = \begin{bmatrix}
\dot{\epsilon} & 0 & 0\\
0 & -\dot{\epsilon} & 0\\
0 & 0 & 0
\end{bmatrix}
\end{equation}
Using vector spherical harmonics, the functional forms of $p'$ and $v_{i}^{H}$ are:
\begin{equation} \label{eq:10}
p' = a E_{jk} \frac{x_j x_k}{r^5}
   \quad\text{and}\quad 
v_{i}^{H} = b E_{ij} \frac{x_j}{r^3} + c T_{ijk} E_{jk}
\end{equation}
where $a$, $b$, and $c$ are scalar constants and the third order tensor $T_{ijk}$ is:
\begin{equation} \label{eq:11}
T_{ijk} = \frac{ \delta_{ij} x_{k} + \delta_{ik} x_{j} + \delta_{jk} x_{i}}{r^5} - \frac{5 x_{i} x_{j} x_{k}}{r^7}
\end{equation}
Satisfying the continuity equation and the no-slip boundary condition on the surface of the sphere, we determine the constants $a$, $b$, and $c$, resulting in an expression for the total fluid velocity for the case of a sphere held fixed in planar extensional flow:
\begin{multline} \label{eq:12}
v_{i}^{total} = v_{i}^{\infty} - \frac{5R^3}{2} \frac{C_i C_j C_k}{C^5} E_{jk}\\
- \frac{R^5}{2} \left[\frac{\delta_{ij} C_{k} + \delta_{ik} C_{j} + \delta_{jk} C_{i}}{C^5} - \frac{5 C_{i} C_{j} C _{k}}{C^7} \right] E_{jk}
\end{multline}
Simplifying the above equation, we obtain two additional terms contributing to the disturbance velocity on particle 1:
\begin{equation} \label{eq:13}
v_{i\:(stresslet,\:pot.\:quad.)}^{D,1} = \left(-\frac{5\dot{\epsilon}R^3}{2C^2} + \frac{3\dot{\epsilon}R^5}{2C^4} \right) \hat{C}_i
\end{equation}
where the subscript ``$pot.\:quad.$" stands for potential quadrupole. Using Eqs. (\ref{eq:7}) and (\ref{eq:13}), the total disturbance velocity acting on particle 1 is:
\begin{align} \label{eq:14}
v_{i}^{D,1} = v_{i\:(Stokeslet)}^{D,1} + v_{i\:(stresslet,\:pot.\:quad.)}^{D,1}\nonumber \\
= \left( \frac{F_{cap}}{4\pi\eta_{c} C} -\frac{5\dot{\epsilon}R^3}{2C^2} + \frac{3\dot{\epsilon}R^5}{2C^4} \right) \hat{C}_i
\end{align}
Similarly, the disturbance velocity acting on particle 2 can be determined using an analogous set of arguments as above. 
 
The force balance on particle 1 becomes:
\begin{multline} \label{eq:16}
6\pi R\eta_c(\dot{\epsilon}y_i + v_{i}^{D,1}) - 2\pi R\sigma \cos\theta\left[1 - \frac{1}{\sqrt{1+\frac{2V}{\pi RD^2}}} \right] \hat{C}_i \\
-\left[ \frac{6\pi\eta_{c} R^2 }{\sqrt{\frac{V}{\pi R} + D^2}}+ \frac{6\pi R^2 }{D}\left(\sqrt{\frac{V}{\pi R}+ D^2}-D\right) \right.\\
\left. \times \frac{\left(\eta_{d} \sqrt{\frac{V}{\pi R}+ D^2} + D(\eta_{c}-\eta_{d})\right)}{\frac{V}{\pi R}+ D^2}\right] \frac{dD}{dt} \hat{C}_i=0
\end{multline}

Similarly, a force balance on particle 2 gives:
\begin{multline} \label{eq:17}
6\pi R\eta_c(\dot{\epsilon}z_i + v_{i}^{D,2}) + 2\pi R\sigma \cos\theta\left[1 - \frac{1}{\sqrt{1+\frac{2V}{\pi RD^2}}} \right] \hat{C}_i \\
+\left[ \frac{6\pi\eta_{c} R^2 }{\sqrt{\frac{V}{\pi R} + D^2}}+ \frac{6\pi R^2 }{D}\left(\sqrt{\frac{V}{\pi R}+ D^2}-D\right) \right.\\
\left. \times \frac{\left(\eta_{d} \sqrt{\frac{V}{\pi R}+ D^2} + D(\eta_{c}-\eta_{d})\right)}{\frac{V}{\pi R}+ D^2}\right] \frac{dD}{dt} \hat{C}_i=0
\end{multline}
Equation (\ref{eq:16}) is subtracted from (\ref{eq:17}), allowing the force balance to be expressed in terms of the vector connecting the bead centers $C_i$. We assume that the two-particle cluster is aligned along the extensional flow axis ($x_1$-axis), which simplifies the force balance to a scalar equation along the $x_1$-axis. This equation is simplified and non-dimensionalized to obtain a first-order differential equation for the dimensionless interparticle separation between bead surfaces $D^*(t^*)$:
\begin{multline} \label{eq:18}
\frac{dD^*}{dt^*} = \frac{Ca \left(3 C^* - \frac{15}{C^{*2}} + \frac{9}{C^{*4}} \right)}{\frac{6}{K^*} + \frac{6(K^*-D^*)(\eta_rK^*+D^*(1-\eta_r))}{D^*K^{*2}}}\\
+\frac{\cos \theta \left(\frac{3}{C^*}-2\right)\left[1 - \frac{1}{\sqrt{1+\frac{2V^*}{\pi D^{*2}}}} \right]}{\frac{6}{K^*} + \frac{6(K^*-D^*)(\eta_rK^*+D^*(1-\eta_r))}{D^*K^{*2}}}
\end{multline}
where the superscript $*$ denotes a dimensionless quantity. Equation (\ref{eq:18}) is non-dimensionalized using the particle radius $R$ and the characteristic time scale $t_c=R\eta_c/\sigma$. The capillary number $Ca=\dot{\epsilon}R\eta_c/\sigma$ is the ratio of the viscous to surface tension forces. The relative viscosity is $\eta_r=\eta_d/\eta_c= 1.9$ for our experimental conditions. The quantity $K^*$ is defined as $K^*=\sqrt{\frac{V^*}{\pi}+D^{*2}}$ and the dimensionless center-to-center distance is $C^*=D^*+2$, where $C=D+2R$. Equation (\ref{eq:18}) is considered for interparticle separations $D/R < 0.5$.

\subsubsection*{Far-field Model}
The far-field model is considered for relatively large interparticle separations ($D/R \ge 0.5$). Here, lubrication forces are not included in the viscous force term $F^{vis,j}_i$. However, the net hydrodynamic drag force is modified to include the center-of-mass bead velocity:
\begin{equation} \label{eq:19}
F^{drag,1}_i=6\pi R\eta_{c}(\dot{\epsilon}y_i - \dot{y_i})
\end{equation}
such that the net viscous force $F^{vis,j}_i$ is now given by $F^{drag,j}_i$ in Eq. (\ref{eq:forcebal}). The capillary force and the disturbance velocity terms remain the same as the near-field model, and the force balance on particle 1 is written as:
\begin{multline} \label{eq:21}
6\pi R\eta_c(\dot{\epsilon}y_i + v_{i}^{D,1}- \dot{y_i})\\
- 2\pi R\sigma \cos\theta\left[1 - \frac{1}{\sqrt{1+\frac{2V}{\pi RD^2}}} \right] \hat{C}_i=0
\end{multline}
Similarly for particle 2:
\begin{multline} \label{eq:22}
6\pi R\eta_c(\dot{\epsilon}z_i + v_{i}^{D,2} - \dot{z_i})\\
+ 2\pi R\sigma \cos\theta\left[1 - \frac{1}{\sqrt{1+\frac{2V}{\pi RD^2}}} \right] \hat{C}_i=0
\end{multline}
Subtracting equation (\ref{eq:21}) from (\ref{eq:22}), simplifying and non-dimensionalizing, we obtain a dimensionless first-order differential equation for the interparticle separation $D^*(t^*)$:
\begin{multline} \label{eq:23}
\frac{dD^*}{dt^*} = Ca \left( C^* - \frac{5}{C^{*2}} + \frac{3}{C^{*4}} \right)\\
+ \cos\theta\left(\frac{1}{C^*} - \frac{2}{3}\right)\left[1 - \frac{1}{\sqrt{1+\frac{2V^*}{\pi D^{*2}}}} \right]
\end{multline}
Thus, from Eqs. (\ref{eq:18}) and (\ref{eq:23}), we obtain:
\begin{multline} \label{eq:24}
    \frac{dD^*}{dt^*}= 
\begin{cases}
    \frac{Ca \left(3 C^* - \frac{15}{C^{*2}} + \frac{9}{C^{*4}} \right)}{\frac{6}{K^*} + \frac{6(K^*-D^*)(\eta_rK^*+D^*(1-\eta_r))}{D^*K^{*2}}}\\
    +\frac{\cos \theta \left(\frac{3}{C^*}-2\right)\left[1 - \frac{1}{\sqrt{1+\frac{2V^*}{\pi D^{*2}}}} \right]}{\frac{6}{K^*} + \frac{6(K^*-D^*)(\eta_rK^*+D^*(1-\eta_r))}{D^*K^{*2}}},\;\;\;\;\;\;\;\;\;\;\;\;\;  \frac{D}{R}<0.5\\
    \\
    Ca \left( C^* - \frac{5}{C^{*2}} + \frac{3}{C^{*4}} \right)\\
    + \cos\theta\left(\frac{1}{C^*} - \frac{2}{3}\right)\left[1 - \frac{1}{\sqrt{1+\frac{2V^*}{\pi D^{*2}}}} \right],\ \frac{D}{R}\geq 0.5
\end{cases}
\end{multline}
Note that $C^*=D^*+2$. Additionally, from equation (\ref{eq:23}), in the limit of large $D^*$, we recover the limit of freely suspended individual particles (such that particles follow fluid streamlines) where particles are assumed to flow freely in planar extensional flow in the absence of any additional forces, {\em i.e.}, $dC^* / dt^*=C^*$.

\subsubsection*{Predicting a Critical Capillary Number $Ca_{crit}$ for Cluster Breakup}
The analytical model developed above for a two-particle liquid-bound cluster breakup can be used to predict the critical capillary number $Ca_{crit}$ for cluster breakup in planar extensional flow. The interplay between various forces during the initial stages of deformation plays a major role in cluster separation. We therefore consider the near-field model (equation (\ref{eq:18})) as given above for this discussion. Equation (\ref{eq:18}) can be written in a compact form as follows:
\begin{equation} \label{eq:Cacrit1}
\frac{dD^*}{dt^*} = \frac{K^*}{6} \left[Ca\cdot f(C^*) + \cos{\theta} \left(\frac{3}{C^*} - 2 \right) g\left(\frac{V^*}{D^{*2}} \right) \right]\frac{1}{h} 
\end{equation}
where $f$, $g$, and $h$ are functions of system parameters. In particular, $f$ is a function of $C^*$ defined as:
\begin{equation}
f(C^*) = 3 C^* - \frac{15}{C^{*2}} + \frac{9}{C^{*4}}
\end{equation}
and $g$ is a function defined as:
\begin{equation}
g\left(\frac{V^*}{D^{*2}} \right) = \left[1 - \frac{1}{\sqrt{1+\frac{2V^*}{\pi D^{*2}}}} \right]
\end{equation}
Finally, $h$ is defined as:
\begin{equation} \label{eq:Cacrit2}
h = 1 + \left(1 - \frac{D^*}{K^*}\right)\left(\eta_r \frac{K^*}{D^*} + (1-\eta_r)  \right)
\end{equation}
Given an initial condition of near contact between particles, cluster separation requires $dD^* / dt^* > 0$. Because $K^* > D^*$ and $h > 0$, the condition for breakup therefore requires:  
\begin{equation} \label{eq:Cacrit3}
Ca\cdot f(C^*) + \cos{\theta} \left(\frac{3}{C^*} - 2 \right) g\left(\frac{V^*}{D^{*2}} \right) > 0
\end{equation}
At time $t=0$, the initial smallest separation between particle surfaces $D_0$ is defined to be $D_0$ = 3 $\mu$m (see Section \ref{sec:results}), which yields $D^*=0.06$. Hence $C^*=2.06$ and $f(2.06)=3.15$. Further, liquid bridge volumes for our experiments typically lie in the range of 1500-2000 $\mu$m$^3$, thus $V^*=V/R^3$ is of the order 0.01. Assuming a value of $V^*=0.01$, $g\left(\frac{V^*}{D^{*2}} \right) \approx 0.4$, which simplifies equation (\ref{eq:Cacrit3}) to:
\begin{equation} \label{eq:Cacrit4}
Ca_{crit} \approx \frac{\cos{\theta}}{14.3}
\end{equation}
Substituting the average value of $\theta=55^o$ for our experiments (see Section \ref{sec:methods}), we obtain a critical capillary number value of $Ca_{crit} = 0.04$, which is in reasonable agreement with the $Ca$ values observed in our experiments (Section \ref{sec:results}). Thus, equation (\ref{eq:Cacrit3}) provides a simple prediction of an approximate $Ca_{crit}$ for a capillary suspension breakup in extensional flow.

\subsection*{Cluster Relaxation}
In this section, we develop a model for cluster relaxation dynamics under zero-flow conditions. In our experiments, a two-particle cluster is trapped near the stagnation point, subjected to extensional flow for a finite time, and the flow is then switched off before cluster breakup occurs. The cluster is then allowed to relax under zero-flow conditions, where the liquid bridge draws the particles together due to capillary forces. Similar to the cluster breakup model, we again develop a near-field and far-field model for the problem. At the beginning of the experiment, the two-particle cluster is significantly deformed and the particles are far apart; hence, lubrication forces are not included in the far-field model. However, after the cluster relaxes due to the attractive capillary force, the particles eventually experience a repulsive lubrication force upon close approach due to the thin liquid film in the gap between particles. 

\subsubsection*{Far-field Model}
In the far-field model, lubrication forces are not included in the viscous force $F^{vis,j}_i$ due to the relatively large interparticle separation. In addition, cluster relaxation occurs under zero-flow conditions. Therefore, the net viscous force arises due to the hydrodynamic drag force on a particle due to its center-of-mass velocity and the disturbance velocity due to hydrodynamic interactions. The drag force on particle 1 is given by:
\begin{equation} \label{eq:25}
F^{drag,1}_i = 6\pi R\eta_{c} (v^{D,1}_i - \dot{y}_i)
\end{equation}
The capillary force on particle 1 due to the liquid bridge (Eq. (\ref{eq:2})) is the same as in the case of cluster breakup. In addition, the disturbance velocity only includes the Stokeslet contribution because of the zero-flow conditions:
\begin{equation} \label{eq:26}
v_{i}^{D,1} = \frac{F_{cap}}{4\pi\eta_{c} C} \hat{C}_i
\end{equation}
For cluster relaxation, we account for excluded volume interactions between the spheres. Here, we introduce the Weeks-Chandler-Anderson (WCA) potential which is the repulsive part of the Lennard-Jones potential. We found that the repulsive potential is needed to capture the physics of near-sphere interaction more accurately. In addition, owing to the polydispersity in the particle size in experiments, perfect quantitative agreement is not expected between experiments and the analytical model. The WCA potential can be defined in terms of the Lennard-Jones potential as follows:
\begin{equation} \label{eq:27}
    V_{WCA}(r)= 
\begin{cases}
    V_{LJ}(r)+\epsilon_{LJ},& \text{if } r<2^{1/6}(C_0)\\
    0,              & \text{if } r\geq 2^{1/6}(C_0)
\end{cases}
\end{equation}
where the Lennard-Jones potential is given by:
\begin{equation} \label{eq:27b}
V_{LJ}=4\epsilon_{LJ}\left[\left(\frac{C_0}{C}\right)^{12} - \left(\frac{C_0}{C}\right)^{6} \right]
\end{equation}
Here, $\epsilon_{LJ}$ is the depth of the potential well and is chosen as the characteristic surface energy ($\sigma R^2$) for our system, $C_0$ is the distance at which the potential energy is zero, and $C$ is the distance between the particle centers. Differentiating equation (\ref{eq:27}), we obtain the repulsive force from the WCA potential on particle 1:
\begin{equation} \label{eq:28}
F^{WCA,1}_{i}=
\begin{cases}
    \frac{4\epsilon_{LJ}}{C}\left[12\left(\frac{C_0}{C}\right)^{12} - 6\left(\frac{C_0}{C}\right)^6\right]\hat{C}_i,& \text{if } C<2^{1/6}C_0\\
    0,              & \text{if } C\geq2^{1/6}C_0
\end{cases}
\end{equation}
The force balance on particle 1 can now be written as: 
\begin{multline} \label{eq:29}
6\pi R\eta_c(v_{i}^{D,1} - \dot{y_i}) \\- 2\pi R\sigma \cos\theta\left[1 - \frac{1}{\sqrt{1+\frac{2V}{\pi RD^2}}} \right] \hat{C}_i + F^{WCA,1}_{i}=0
\end{multline}
Similarly for particle 2:
\begin{multline} \label{eq:30}
6\pi R\eta_c(v_{i}^{D,2} - \dot{z_i}) \\+ 2\pi R\sigma \cos\theta\left[1 - \frac{1}{\sqrt{1+\frac{2V}{\pi RD^2}}} \right] \hat{C}_i + F^{WCA,2}_{i}=0
\end{multline}
Subtracting equation (\ref{eq:29}) from (\ref{eq:30}), simplifying and non-dimensionalizing, we obtain a dimensionless first-order differential equation for the interparticle spacing $D^*$ describing cluster relaxation:
\begin{equation} \label{eq:31}
\frac{dD^*}{dt^*} = \cos\theta\left(\frac{1}{C^*}-\frac{2}{3}\right)\left[1 - \frac{1}{\sqrt{1+\frac{2V^*}{\pi D^{*2}}}} \right] + \frac{F^*_{WCA}}{3}
\end{equation}
where $F^*_{WCA}$ is the scalar dimensionless force from the WCA potential given as:
\begin{equation} \label{eq:32}
F^*_{WCA}=
\begin{cases}
    \frac{\epsilon_c}{C^*}\left[12\left(\frac{C^*_0}{C^*}\right)^{12} - 6\left(\frac{C^*_0}{C^*}\right)^6\right],& \text{if } C^*<2^{1/6}(C^*_0)\\
    0,              & \text{if } C^*\geq 2^{1/6}(C^*_0)
\end{cases}
\end{equation}
where $\epsilon_c$ is the dimensionless dispersion energy defined as $\epsilon_c = 4\epsilon_{LJ}/\pi\sigma R^2$ and is taken as a fitting parameter in the model.

As the cluster begins to relax and the particles approach each other, lubrication forces will begin to play a role in the process. In our experiments, we observe a marked decrease in the rate of cluster relaxation at a finite interparticle separation, which occurs when lubrication forces begin to play a role in the relaxation process (see Figure \ref{fig:relaxnum}). Based on our experimental data, we define the transition point from the far-field to the near-field model when $\approx 75\%$ of the relaxation has occurred. Therefore, the far-field model is assumed to transition to the near-field model at a dimensionless interparticle spacing of $D/R = 0.2$ during cluster relaxation.

\subsubsection*{Near-field Model}
In the near-field case, the particles experience a repulsive lubrication force due to the thin liquid film. As discussed above, the lubrication force given by Eq. (\ref{eq:3}) accounts for the particle approach velocity (and hence viscous drag due to particle center-of-mass motion). Therefore, the hydrodynamic drag force only accounts for the disturbance velocity at particle 1:
\begin{equation} \label{eq:33}
F^{drag,1}_i = 6\pi R\eta_{c}v^{D,1}_i
\end{equation}
Combining the expressions for the capillary force and excluded volume forces from Eqs. (\ref{eq:2}) and (\ref{eq:28}) respectively, the force balance on particle 1 is given as:
\begin{multline}\label{eq:34}
6\pi R\eta_c v_{i}^{D,1} - 2\pi R\sigma \cos\theta\left[1 - \frac{1}{\sqrt{1+\frac{2V}{\pi RD^2}}} \right] \hat{C}_i + F^{WCA,1}_{i}\\
-\left[ \frac{6\pi\eta_{c} R^2 }{\sqrt{\frac{V}{\pi R} + D^2}}+ \frac{6\pi R^2 }{D}\left(\sqrt{\frac{V}{\pi R}+ D^2}-D\right) \right.\\
\left. \times \frac{\left(\eta_{d} \sqrt{\frac{V}{\pi R}+ D^2} + D(\eta_{c}-\eta_{d})\right)}{\frac{V}{\pi R}+ D^2}\right] \frac{dD}{dt} \hat{C}_i=0
\end{multline}
Similarly a force balance on particle 2 yields:
\begin{multline}\label{eq:35}
6\pi R\eta_c v_{i}^{D,2} + 2\pi R\sigma \cos\theta\left[1 - \frac{1}{\sqrt{1+\frac{2V}{\pi RD^2}}} \right] \hat{C}_i + F^{WCA,2}_{i}\\
+\left[ \frac{6\pi\eta_{c} R^2 }{\sqrt{\frac{V}{\pi R} + D^2}}+ \frac{6\pi R^2 }{D}\left(\sqrt{\frac{V}{\pi R}+ D^2}-D\right) \right.\\
\left. \times \frac{\left(\eta_{d} \sqrt{\frac{V}{\pi R}+ D^2} + D(\eta_{c}-\eta_{d})\right)}{\frac{V}{\pi R}+ D^2}\right] \frac{dD}{dt} \hat{C}_i=0
\end{multline}

\begin{figure*}[t]
\includegraphics[width=0.75\textwidth]{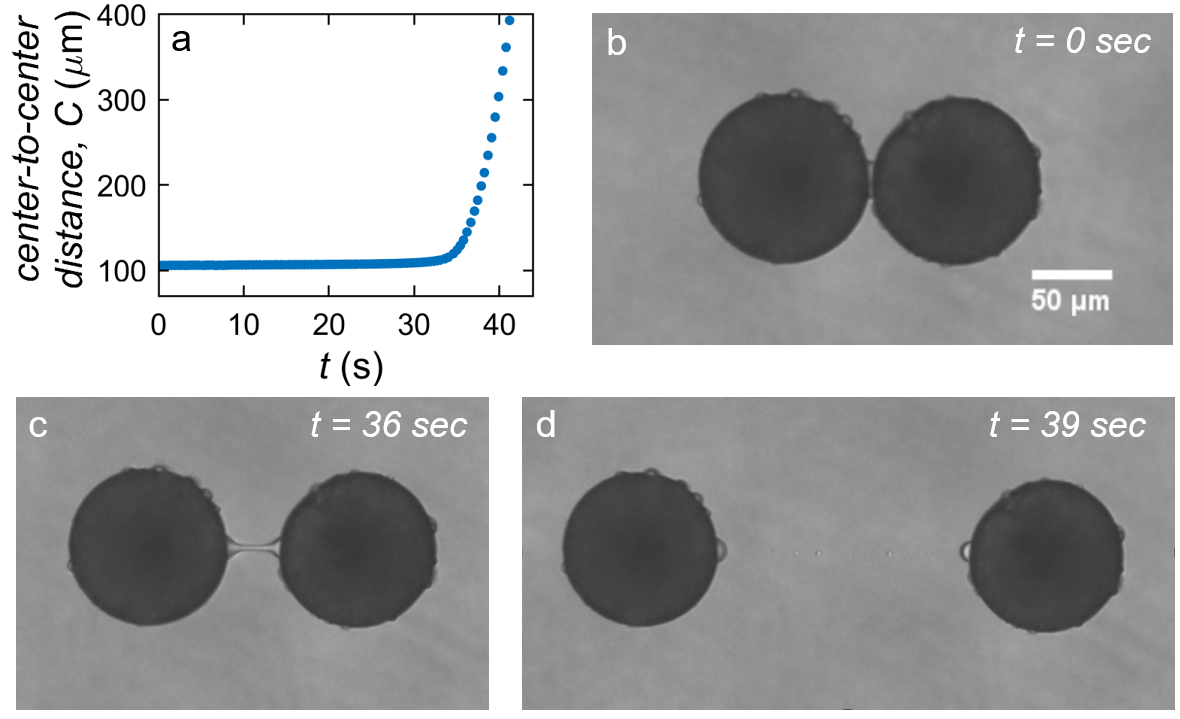}
\caption{\label{fig:breakupsnap} Experiments on two-particle cluster breakup in extensional flow (a) Center-to-center distance $C$ plotted as a function of time for a two-particle cluster breakup for $Ca = 0.035$. (b) Snapshot of the cluster at $t = 0$ sec. For consistency, time $t=0$ in all experiments is defined as the time at which interparticle separation is $D=3\: \mu$m. (c) Snapshot of the cluster at $t = 36$ sec. Here, the cluster has started to separate and the meniscus is stretched into a thin thread. (d) Snapshot of the cluster at $t = 39$ sec, which occurs after the cluster is completely broken up.}
\end{figure*}

Following a similar procedure as explained above, we obtain a dimensionless first-order differential equation for interparticle separation $D^*$ for the near-field case of cluster relaxation:
\begin{equation} \label{eq:36}
\frac{dD^*}{dt^*} = \frac{\cos \theta \left(\frac{3}{C^*}-2\right)\left[1 - \frac{1}{\sqrt{1+\frac{2V^*}{\pi D^{*2}}}} \right] + F^*_{WCA}}{\frac{6}{K^*} + \frac{6(K^*-D^*)(\eta_rK^*+D^*(1-\eta_r)))}{D^*K^{*2}}}
\end{equation}
Taken together, Eqs. (\ref{eq:31}) and (\ref{eq:36}) provide a model for cluster relaxation in terms of a dimensionless interparticle separation $D^*(t^*)$:
\begin{equation} \label{eq:37}
    \frac{dD^*}{dt^*}= 
\begin{cases}
    \cos\theta\left(\frac{1}{C^*}-\frac{2}{3}\right)\left[1 - \frac{1}{\sqrt{1+\frac{2V^*}{\pi D^{*2}}}} \right] + \frac{F^*_{WCA}}{3},\; \frac{D}{R}>0.2\\
    \\
    \frac{\cos \theta \left(\frac{3}{C^*}-2\right)\left[1 - \frac{1}{\sqrt{1+\frac{2V^*}{\pi D^{*2}}}} \right] + F^*_{WCA}}{\frac{6}{K^*} + \frac{6(K^*-D^*)(\eta_rK^*+D^*(1-\eta_r)))}{D^*K^{*2}}},\;\;\;\;\;\;\;\;\;\;\;\;\;\;\;\: \frac{D}{R}\leq 0.2
\end{cases}
\end{equation}
Note that $C^*=D^*+2$. In the limit of small $D^*$, the near-field equations for both breakup and relaxation predict exponential separation or approach of particles, respectively.

\section{\label{sec:results}Results and Discussion}

\begin{figure*}[t]
\includegraphics[width=0.9\textwidth]{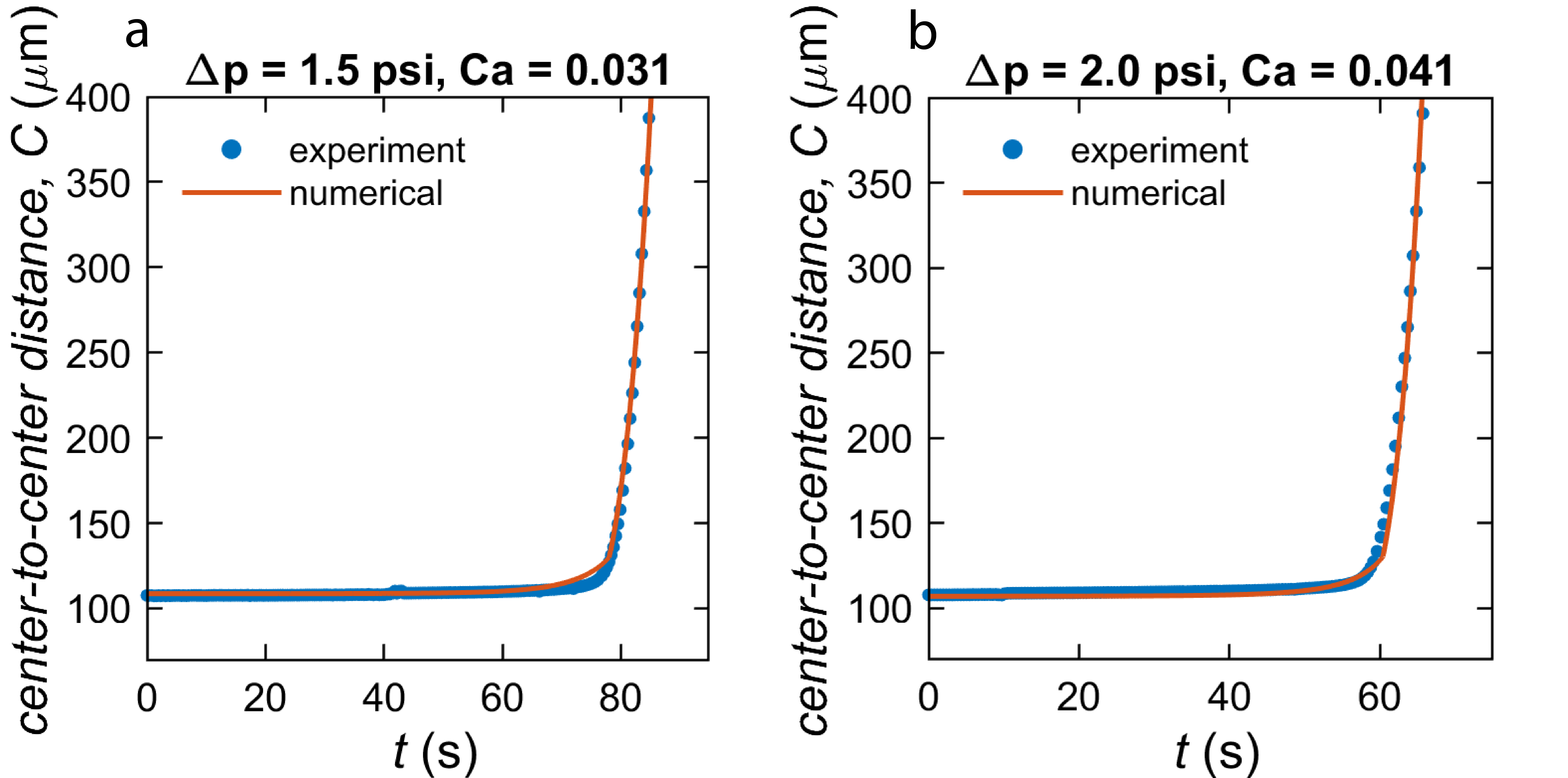}
\caption{\label{fig:breaknum} Center-to-center distance $C$ between particles plotted as a function of time for cluster breakup for two different $Ca$, (a) $Ca = 0.031$ and (b) $Ca = 0.041$. The blue scatter points represent the experimental data which is obtained through image analysis at each frame, and the solid orange line corresponds to the analytical model for cluster breakup where a first order differential equation is solved numerically using RK4, yielding good agreement with the experiments.}
\end{figure*}

\subsection*{Cluster Breakup}
We began by experimentally observing cluster breakup for two-particle clusters using a Stokes trap. In this experiment, we measured the cluster breakup time as a function of strain rate $\dot{\epsilon}$ or capillary number $Ca$, defined as:
\begin{equation} \label{eq:38}
Ca=\frac{\dot{\epsilon}R\eta_c}{\sigma}
\end{equation}
Two-particle clusters are introduced into the cross-slot device. A target cluster is trapped and positioned to be located near the center of the cross-slot, and the flow is switched off. Two-particle clusters near the center plane of the flow device (in the $x_3$-direction) were considered for this work. The experiment begins by turning on the planar extensional flow while confining the cluster near the stagnation point using the Stokes trap. For consistency, time $t=0$ in all experiments is defined at the time at which interparticle separation is $D$ = 3 $\mu$m using image analysis (see below). After the extensional flow is turned on, the two-particle cluster reorients and aligns along the direction of the principal axis of extension. The cluster is then subjected to extensional flow until the meniscus stretches and ultimately ruptures, leading to cluster breakup (supplementary material, Movie 2). The transient cluster breakup process is shown in Figure \ref{fig:breakupsnap}(a), where the center-to-center distance between particles in a two-particle cluster is shown for $Ca$ = 0.035. Figures \ref{fig:breakupsnap}(b),(c), and (d) show snapshots of the cluster at different times during the breakup process. Figure \ref{fig:breakupsnap}(b) shows the cluster at $t=0$ sec, when the cluster has already reoriented. Figure \ref{fig:breakupsnap}(c) shows the cluster at $t=36$ sec, when the meniscus has stretched into a thin thread immediately prior to rupture. Figure \ref{fig:breakupsnap}(d) shows an image of the system at time $t=39$ sec, which occurs after the cluster has broken up and the two particles are separated in flow. After the meniscus ruptures and the cluster has broken up, the particles rapidly separate as a function of time in flow. 

Images of cluster breakup events from experiments are analyzed using a MATLAB program. The image is first binarized and impurities surrounding the cluster are removed. To identify and track individual particles in the two-particle cluster, a watershed segmentation algorithm is used \cite{vincent1991watersheds, preim2013visual}. The centroid positions of each particle are then determined, which enables calculation of the center-to-center distance $C$. Breakup experiments are performed until the meniscus is completely broken, which occurs when $C$ is approximately 4 times larger than its initial value i.e. roughly 4 times the particle diameter. Using this method, we quantify the interparticle separation as a function of time during a cluster breakup experiment, as shown in Figure \ref{fig:breakupsnap}(a).

Results from cluster breakup experiments are compared to the analytical model developed in Section \ref{sec:model}. The first-order differential equation for cluster breakup (given by Eq. (\ref{eq:24})) is solved numerically using the fourth-order Runge-Kutta method (RK4). After solving equation (\ref{eq:24}) for the dimensionless interparticle separation $D^*$, the quantities are converted back to their dimensional forms using the relation $t^* = t / t_c$, where the characteristic time scale is $t_c=R\eta_c/\sigma$, enabling direct comparison with experiments. Due to limitations in spatial resolution in optical imaging experiments, the initial value of $D$ at time $t=0$ may vary slightly ($D_0$ = 3 $\pm$ 1.6 $\mu$m) between different experiments. Therefore, a non-linear least squares analysis is used to fit the experimental data to the numerical solution by treating $D_0$ as a fitting parameter. We note that for the cluster breakup model, $D_0$ is the {\em only} fitting parameter used when comparing to experimental data. 

Figure \ref{fig:breaknum} provides a comparison between cluster breakup experiments and the analytical model, generally showing good agreement for two different flow strengths ($Ca=0.031$ and $Ca=0.041$). Using the non-linear least squares analysis, the initial values $D_0$ for the data shown in Figures \ref{fig:breaknum}(a) and (b) are $D_0$ = 4.6 $\mu$m and 3.0 $\mu$m, respectively. The contact angle $\theta$ is measured using image analysis software (ImageJ), and the meniscus volume $V$ is determined using image analysis. Our results show that the meniscus volume $V$ typically lies in the range 1500-2000 $\mu$m$^3$. In addition, the model shows that lubrication and capillary forces generally dominate for small particle separations ($D/R < 0.5$). As $D$ gradually increases, $F^*_{drag}/F^*_{cap}$ becomes greater than 1 and starts increasing rapidly for $D/R > 0.5$. For larger particle separations, the hydrodynamic drag force dominates and the meniscus rapidly stretches and thins, eventually resulting in cluster breakup. Overall, the analytical model accurately captures the breakup dynamics of two-particle meniscus bound clusters in extensional flow.

\begin{figure*}[t]
\includegraphics[width=0.9\textwidth]{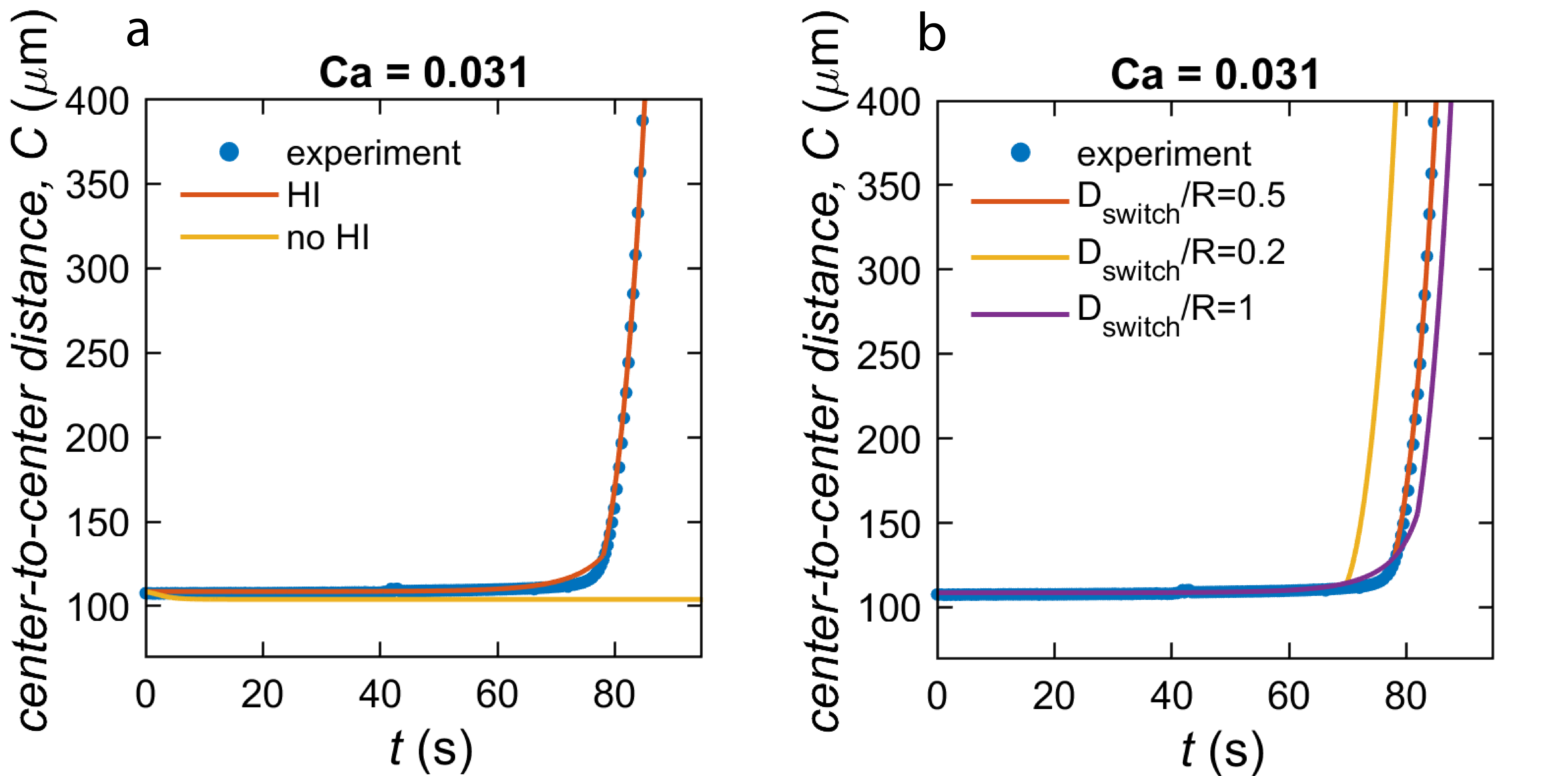}
\caption{\label{fig:HIandDR} Role of interparticle hydrodynamic interactions (HI) and transition between near-field and far-field model on cluster breakup dynamics. (a) Cluster breakup analytical models with and without including HI are shown. For the case with no HI, the model does not predict cluster breakup, and the particles do not separate in flow. (b) Effect of varying the transition point $D_{switch}/R$ at which the model switches from the near-field to the far-field case. }
\end{figure*}

We further examined the role of interparticle hydrodynamic interactions (HI) on the cluster breakup process. Figure \ref{fig:HIandDR}(a) shows a comparison between analytical models for cluster breakup with HI ({\em i.e.}, Stokeslet, stresslet, and potential dipole interactions between beads) and without HI. The model incorporating HI yields good agreement with experiments. However, the model without HI does not predict cluster breakup, and the particles never separate under identical flow conditions. This comparison reveals the sensitivity of the breakup process to small perturbations in the flow around the particles and emphasizes the importance of interparticle HI. We note that the analytical model presented here only considers first-order interactions between beads and does not consider higher order corrections via the method of reflections {\em i.e.,} second-order and higher reflections are not considered here \cite{kimtext}. Nevertheless, higher order corrections are not expected to contribute to significant differences in the general qualitative behavior or outcomes observed here. The next higher order term ($\mathcal{O}(1/C^5)$) in the multipole expansion is $\approx$20 times smaller than the Stokeslet interaction. We also probed the effect of changing the transition point $D_{switch}/R$ when the analytical model switches from the near-field to the far-field case (Section \ref{sec:model}). Figure \ref{fig:HIandDR}(b) shows the comparison for various $D_{switch}/R$; our results show that changing the transition point value of $D_{switch}/R=0.5$ does not qualitatively affect the results. In brief, immediately prior to cluster breakup, the meniscus begins to rapidly stretch and $D$ changes rapidly. Therefore, the transition point $D_{switch}/R$ plays a relatively minor role on the numerical results, as shown in Figure \ref{fig:HIandDR}(b).

\begin{figure*}[t]
\includegraphics[width=0.9\textwidth]{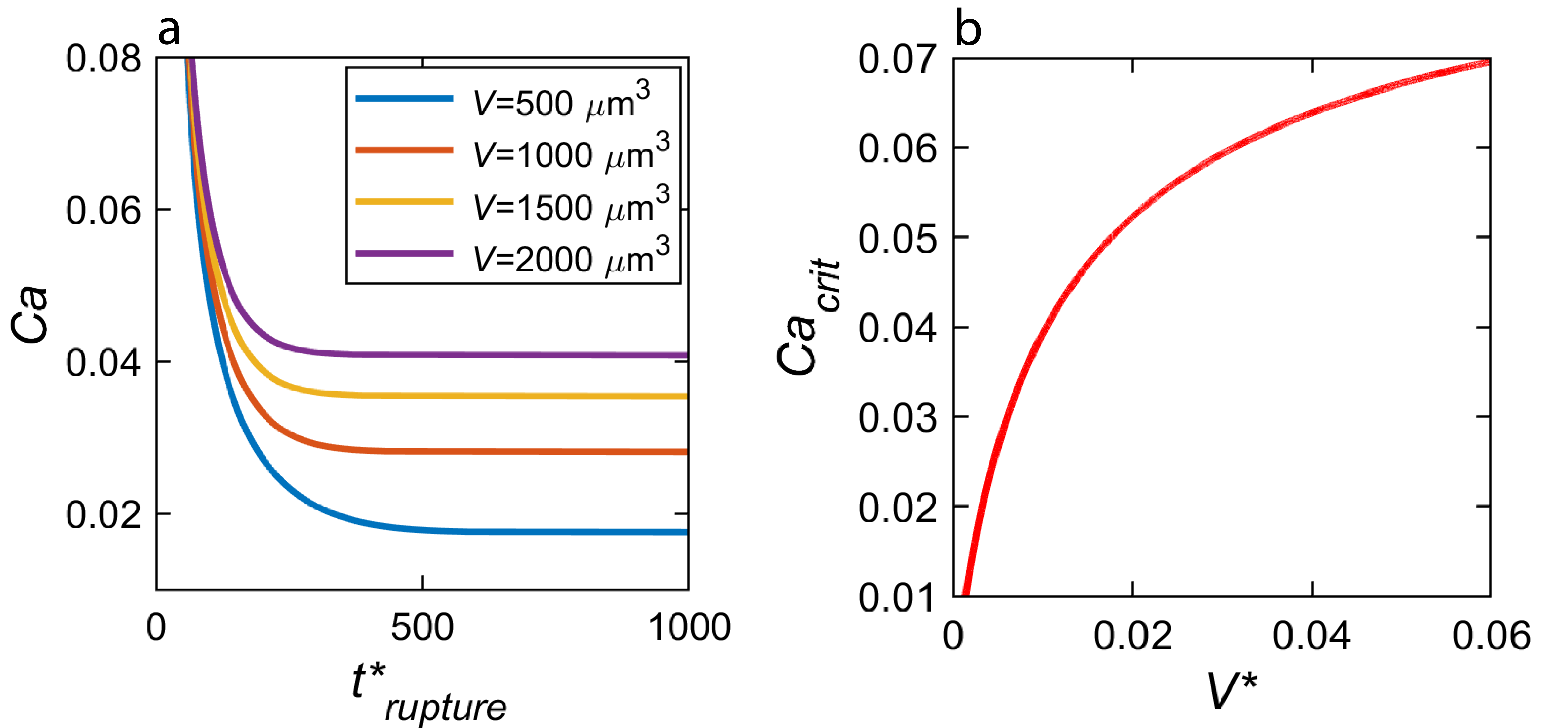}
\caption{\label{fig:Ca_time} The dependence of cluster breakup time on $Ca$ and $V$ and the prediction of a critical capillary number $Ca_{crit}$ for two-particle clusters in extensional flow. (a) Dimensionless breakup time $t^*_{rupture}$ is plotted as a function of $Ca$ for selected values of $V$ using the results from the analytical model. (b) $Ca_{crit}$ is plotted as a function of the dimensionless liquid bridge volume $V^*$.}
\end{figure*}

We next investigated the dependence of cluster breakup time on capillary number $Ca$ and liquid bridge volume $V$. Based on the cluster breakup analytical model in Section \ref{sec:model}, we observe that given a $Ca$ and $V$ value, equation (\ref{eq:24}) can be solved to compute the corresponding cluster breakup time. Figure \ref{fig:Ca_time}(a) shows this variation where the dimensionless cluster breakup time $t^*_{rupture}$ is plotted as a function of $Ca$ for various liquid bridge volumes as indicated in the legend. Similar to the experiments, cluster breakup is considered to occur when $C$ is approximately a factor of 4 larger than the initial value. It can be seen from the figure that below a certain value of $Ca$, $t^*_{rupture}$ increases rapidly and the curve flattens out. This turning point in the plot can be interpreted as the critical capillary number $Ca_{crit}$ for a given liquid bridge volume, below which rupture does not occur. Furthermore, using the expression for predicting $Ca_{crit}$ as derived in Section \ref{sec:model} (equation (\ref{eq:Cacrit3})), we plot $Ca_{crit}$ as a function of dimensionless liquid bridge volume $V^*$, as shown in Figure \ref{fig:Ca_time}(b). Experimentally, we observe that for flow strengths below the $Ca_{crit}$ line in Figure \ref{fig:Ca_time}(b), clusters with liquid bridge volumes within the range indicated generally do not separate. For these calculations, the average contact angle is $\theta=55^o$ and the $D^*=0.06$.

\subsection*{Cluster Relaxation}

We further studied cluster relaxation under zero-flow conditions. The first part of the experimental procedure is similar to the cluster breakup experiment. Here, two-particle clusters are trapped near the stagnation point in the cross-slot device and subjected to planar extensional flow. However, in the cluster relaxation experiment, as the meniscus begins to stretch, the flow is stopped immediately prior to separation, and the cluster is allowed to relax under zero-flow conditions (supplementary material, Movie 3). 

\begin{figure*}[t]
\includegraphics[width=0.8\textwidth]{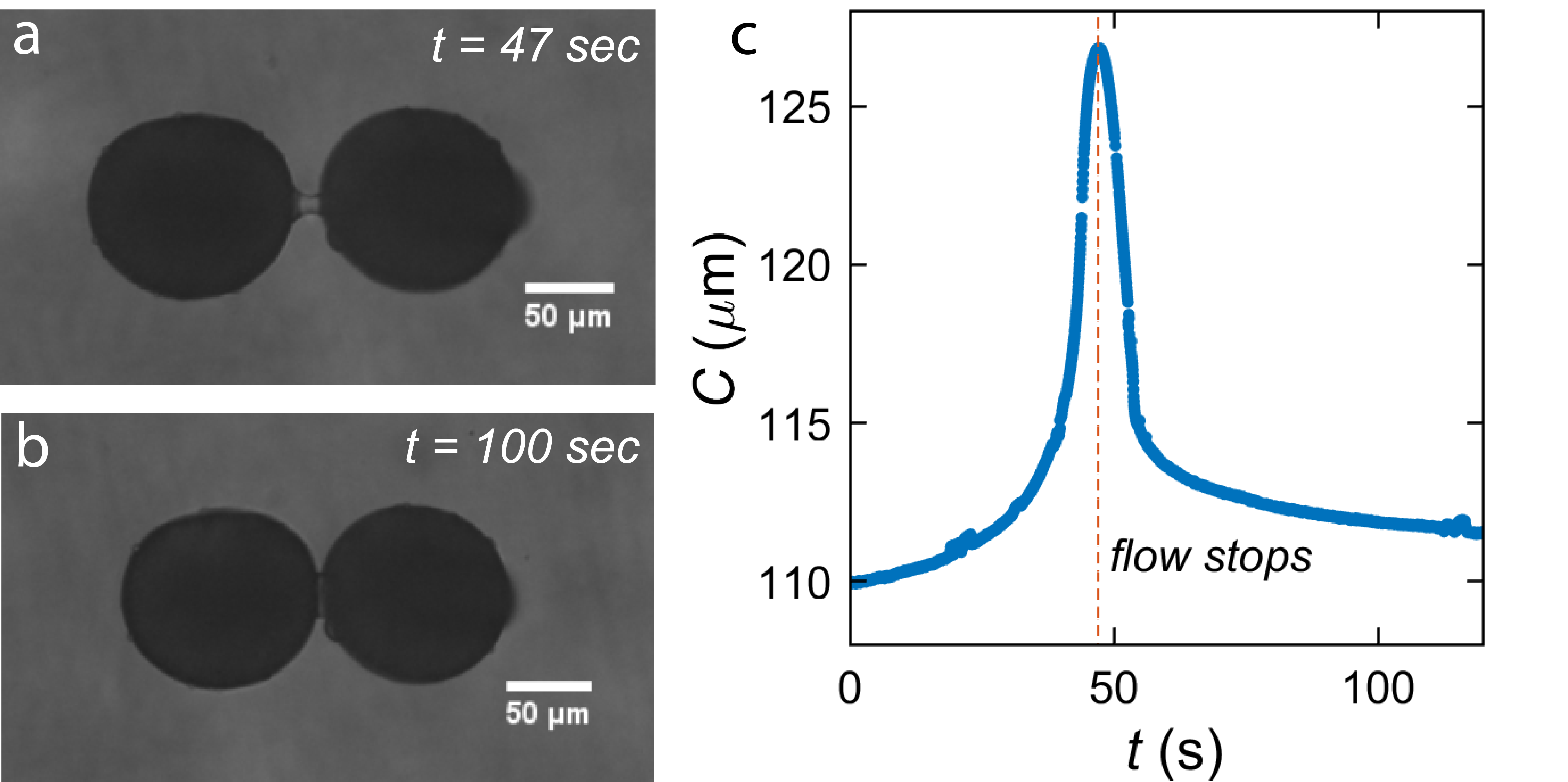}
\caption{\label{fig:relaxexp} Two-particle cluster relaxation under zero-flow conditions. (a) Snapshot of a two-particle cluster at $t = 47$ sec, where the meniscus is stretched to the point immediately prior to rupture and the flow is turned off. Following cessation of flow, the cluster begins to relax, and the particles come closer to each other. (b) Snapshot of the cluster at a later time ($t = 100$ sec), when the cluster has sufficiently relaxed and particle separation is close to the initial separation. (c) Center-to-center distance, $C$ plotted as a function of time for cluster stretching and relaxing. The red dashed line marks the instance when the flow is turned off and the cluster begins to relax ($t = 47$ sec).}
\end{figure*}

\begin{figure}[t]
\includegraphics[width=0.45\textwidth]{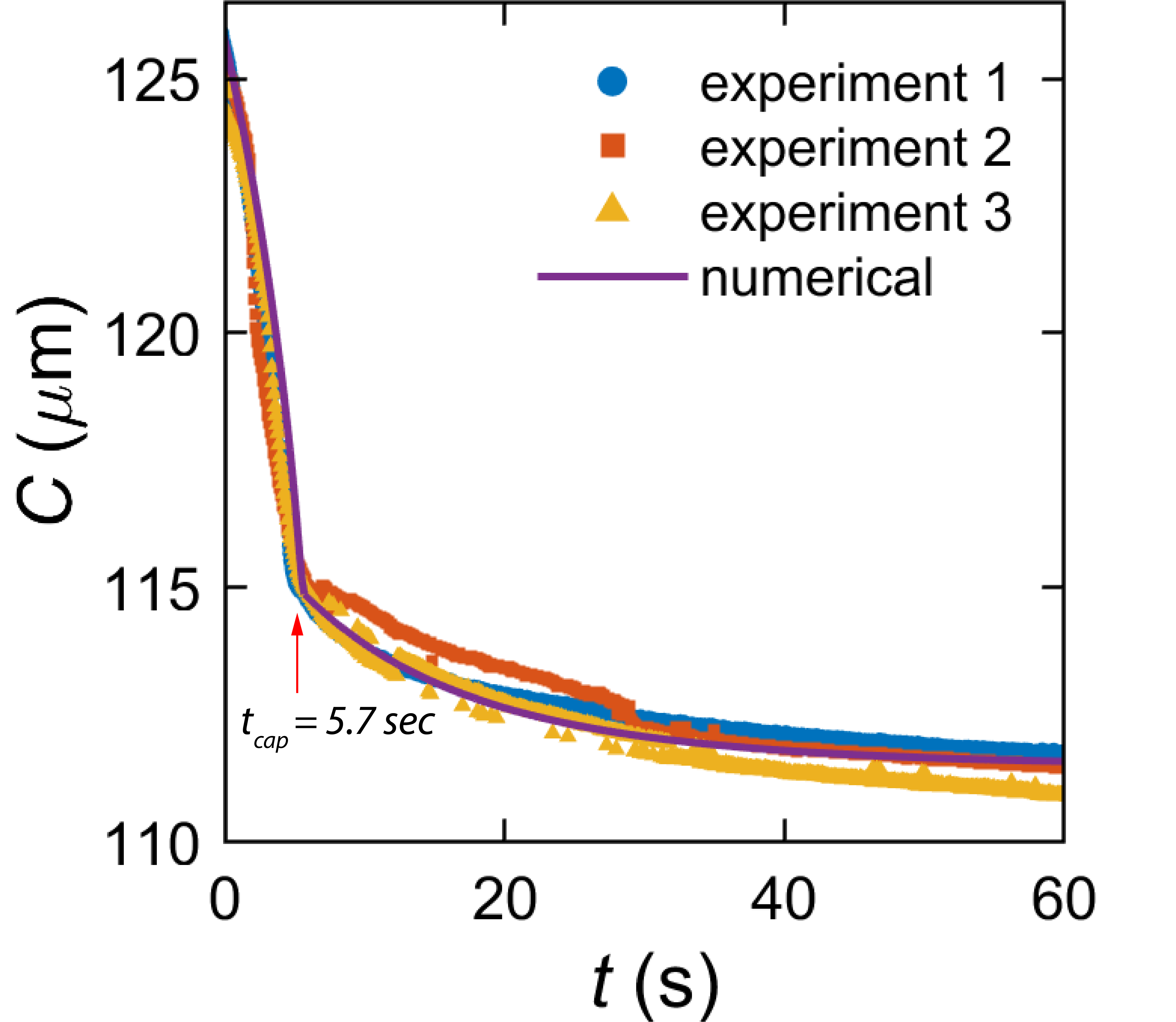}
\caption{\label{fig:relaxnum} Center-to-center distance $C$ plotted as a function of time for cluster relaxation under zero-flow conditions. Experimental data is plotted for three different cluster relaxation experiments and is compared with the analytical model obtained by numerically solving the first-order differential equation for cluster relaxation (Section \ref{sec:model}). Our results show two modes of relaxation: an initial fast mode, dominated by the capillary force, and a second, slower mode dominated by lubrication forces. In solving the analytical model, the initial value for $D$ is chosen to be $D_0$ = 22 $\mu$m to match experiments. The contact angle is $\theta=$ 50$^o$ and the meniscus volume $V \approx$ 2000 $\mu$m$^3$.}
\end{figure}

Figures \ref{fig:relaxexp}(a) and (b) show two different stages of a two-particle cluster during relaxation. Figure \ref{fig:relaxexp}(a) shows a snapshot of the cluster at $t=47$ sec when the flow is stopped prior to cluster breakup. The snapshot in Figure \ref{fig:relaxexp}(b) shows the two-particle cluster at a later time following cessation of flow, after which the meniscus has sufficiently relaxed and the inter-particle distance is close to the initial separation. Figure \ref{fig:relaxexp}(c) shows the interparticle separation $C$ as a function of time for both the deformation and relaxation phase of the two-particle cluster. The red dashed line marks the instant in time when the flow is turned off, which defines the start time of the relaxation experiment. Equation (\ref{eq:37}) is solved numerically to determine $D^*$ as a function of time using the RK4 method, and the results are compared with three different relaxation experiments. As described in Section \ref{sec:model}, the dimensionless dispersion energy $\epsilon_c$ (Eq. (\ref{eq:32})) is used as a fitting parameter, and a non-linear least squares analysis is used to fit the experimental data to the analytical solution. Here, $\epsilon_c = 0.25$, which validates our choice of $\epsilon_{LJ}$ to be of the order of the characteristic surface energy $\sigma R^2$. Generally, the dispersion energy for soft sphere potential is chosen to be of the order of $k_B T$ in literature \cite{koura1991variable, rabani2001interactions, liebetreu2020cluster}. However, the model required a slightly stronger magnitude repulsion to prevent particles from passing into each other. Because the characteristic surface energy for the system is $\sigma R^2 \approx 10^9$ $k_B T$, $\epsilon_{LJ}$ was taken to be of a similar order as $\sigma R^2$. The distance $C_0$ at which the potential energy $V_{LJ}$ is zero (Equations (\ref{eq:27b}, \ref{eq:28})) is $C_0=105$ $\mu$m for the model. The dimensionless quantities are converted back to their dimensional forms for direct comparison with experiments (Figure \ref{fig:relaxnum}). Our results generally show good agreement between experiments and the analytical model for particle cluster relaxation following cessation of extensional flow. Transient relaxation data are plotted until the cluster has relaxed to a final $C$ value within 2\% of the initial $C$ value at time $t=0$ in Figure \ref{fig:relaxexp}(c).

Cluster relaxation experiments clearly show two different regimes of relaxation dynamics. In particular, cluster relaxation begins with a rapid retraction of the liquid bridge that eventually transitions into a second, slower relaxation regime (Figure \ref{fig:relaxnum}). At relatively large particle separations ($D/R > 0.2$), the cluster relaxation process is dominated by the capillary force, leading to rapid relaxation ({\em i.e.,} the far-field case). However, in the second regime ({\em i.e.} the near-field case), lubrication forces begin to play a role, and the particles experience a resisting force from the thin liquid film between the particle surfaces, leading to much slower relaxation. To quantify this behavior, we define a characteristic relaxation time $t_{cap}$ for the capillary relaxation mode (first regime) as the time required for 75-80\% of the relaxation to occur. Using the analytical model, we find $t_{cap}=5.7$ sec (denoted by the red arrow in Figure \ref{fig:relaxnum}), which is only $\approx 9.5\%$ of the total relaxation time. 

\section{\label{sec:conclusion}Conclusion}
In this work, we investigate the dynamics of meniscus bound two-particle clusters in extensional flow using an automated flow-based technique known as the Stokes trap. In all cases, the experimental results are complemented with analytical models. Our results show that during the breakup process, clusters begin by aligning along the principal axis of extension in planar extensional flow, followed by deformation of the liquid bridge and eventual breakup after the meniscus has stretched into a thin thread. The breakup time is affected by several forces and parameters, and the overall breakup process is well described by the analytical model. The interplay between the hydrodynamic drag force, capillary force, lubrication force, and interparticle HI is crucial in determining the breakup dynamics of these clusters. In general the experimental data are in good agreement with results from the analytical model. Using the cluster breakup analytical model, we derive an expression to predict the critical capillary number $Ca_{crit}$ in terms of the contact angle, initial particle separation, and the liquid bridge volume for cluster breakup in planar extensional flow. Cluster breakup time is also quantified as a function of $Ca$ and $V$, and the dimensionless breakup time $t^*_{rupture}$ is plotted against $Ca$ for various values of $V$. Our results show that larger $Ca$ and smaller $V$ values lead to smaller breakup times, whereas smaller $Ca$ and larger $V$ values result in longer breakup times.  The relaxation dynamics of two-particle clusters were further characterized under zero-flow conditions. Our results show that cluster relaxation is a dual-mode process with the first, faster relaxation mode dominated by capillary forces, and a second, slower mode is dominated by the lubrication force. An analytical model is developed for cluster relaxation, and good agreement is obtained between experiments and predictions from the analytical model. Overall, our work provides a new understanding of the dynamics and behavior of cluster breakup and relaxation for freely suspended particle clusters in extensional flow, which can be used to inform the processing of capillary suspensions for improved design of mixing operations. 

\section*{\label{sec:ack}Acknowledgements}
This work was supported by the National Science Foundation under Grant No. CBET-2030537. The authors thank Jovina Vaswani from the University of Pittsburgh for providing materials and useful discussions. The authors also thank Hung Nguyen and Lehan Yao from the University of Illinois Urbana-Champaign for help with image analysis.

\section*{\label{sec:dec}Author Declarations}
\subsection*{Conflict of Interest}
The authors have no conflicts to disclose.

\section*{\label{sec:data}Data Availability}
The data that support the findings of this study are available from the corresponding author upon reasonable request.
\nocite{*}
\bibliography{Chaudhary}

\providecommand{\noopsort}[1]{}\providecommand{\singleletter}[1]{#1}%
\begin{thebibliography}{10}

\bibitem{koos2014capillary}
Erin Koos.
\newblock Capillary suspensions: Particle networks formed through the capillary
  force.
\newblock {\em Current opinion in colloid \& interface science},
  19(6):575--584, 2014.

\bibitem{koos2011capillary}
Erin Koos and Norbert Willenbacher.
\newblock Capillary forces in suspension rheology.
\newblock {\em Science}, 331(6019):897--900, 2011.

\bibitem{schneider2017suppressing}
Monica Schneider, Johannes Maurath, Steffen~B Fischer, Moritz Wei{\ss}, Norbert
  Willenbacher, and Erin Koos.
\newblock Suppressing crack formation in particulate systems by utilizing
  capillary forces.
\newblock {\em ACS applied materials \& interfaces}, 9(12):11095--11105, 2017.

\bibitem{amoabeng2018bulk}
Derrick Amoabeng and Sachin~S Velankar.
\newblock Bulk soldering: Conductive polymer composites filled with copper
  particles and solder.
\newblock {\em Colloids and Surfaces A: Physicochemical and Engineering
  Aspects}, 553:624--632, 2018.

\bibitem{koos2012tuning}
Erin Koos, Julia Johannsmeier, Linda Schwebler, and Norbert Willenbacher.
\newblock Tuning suspension rheology using capillary forces.
\newblock {\em Soft Matter}, 8(24):6620--6628, 2012.

\bibitem{hoffmann2014using}
Susanne Hoffmann, Erin Koos, and Norbert Willenbacher.
\newblock Using capillary bridges to tune stability and flow behavior of food
  suspensions.
\newblock {\em Food Hydrocolloids}, 40:44--52, 2014.

\bibitem{roh20173d}
Sangchul Roh, Dishit~P Parekh, Bhuvnesh Bharti, Simeon~D Stoyanov, and Orlin~D
  Velev.
\newblock 3d printing by multiphase silicone/water capillary inks.
\newblock {\em Advanced Materials}, 29(30):1701554, 2017.

\bibitem{dittmann2013ceramic}
Jens Dittmann, Erin Koos, and Norbert Willenbacher.
\newblock Ceramic capillary suspensions: Novel processing route for macroporous
  ceramic materials.
\newblock {\em Journal of the American Ceramic Society}, 96(2):391--397, 2013.

\bibitem{studart2006processing}
Andre~R Studart, Urs~T Gonzenbach, Elena Tervoort, and Ludwig~J Gauckler.
\newblock Processing routes to macroporous ceramics: a review.
\newblock {\em Journal of the American Ceramic Society}, 89(6):1771--1789,
  2006.

\bibitem{schneider2016highly}
Monica Schneider, Erin Koos, and Norbert Willenbacher.
\newblock Highly conductive, printable pastes from capillary suspensions.
\newblock {\em Scientific reports}, 6(1):31367, 2016.

\bibitem{yang2019magnetorheological}
Jianjian Yang, Zhide Hu, Hua Yan, and Fanghao Niu.
\newblock Magnetorheological suspension with capillary network.
\newblock {\em Journal of Intelligent Material Systems and Structures},
  30(12):1850--1857, 2019.

\bibitem{van1975rheology}
Sheau Van~Kao, Lawrence~E Nielsen, and Christopher~T Hill.
\newblock Rheology of concentrated suspensions of spheres. ii. suspensions
  agglomerated by an immiscible second liquid.
\newblock {\em Journal of Colloid and Interface Science}, 53(3):367--373, 1975.

\bibitem{hauf2018structure}
Katharina Hauf and Erin Koos.
\newblock Structure of capillary suspensions and their versatile applications
  in the creation of smart materials.
\newblock {\em MRS communications}, 8(2):332--342, 2018.

\bibitem{bossler2017influence}
Frank Bossler, Lydia Weyrauch, Robert Schmidt, and Erin Koos.
\newblock Influence of mixing conditions on the rheological properties and
  structure of capillary suspensions.
\newblock {\em Colloids and Surfaces A: Physicochemical and Engineering
  Aspects}, 518:85--97, 2017.

\bibitem{mcfarlane1950adhesion}
Jo~S McFarlane and David Tabor.
\newblock Adhesion of solids and the effect of surface films.
\newblock {\em Proceedings of the Royal Society of London. Series A.
  Mathematical and Physical Sciences}, 202(1069):224--243, 1950.

\bibitem{mason1965liquid}
Geoffrey Mason and WC~Clark.
\newblock Liquid bridges between spheres.
\newblock {\em Chemical Engineering Science}, 20(10):859--866, 1965.

\bibitem{willett2000capillary}
Christopher~D Willett, Michael~J Adams, Simon~A Johnson, and Jonathan~PK
  Seville.
\newblock Capillary bridges between two spherical bodies.
\newblock {\em Langmuir}, 16(24):9396--9405, 2000.

\bibitem{pitois2000liquid}
Olivier Pitois, Pascal Moucheront, and Xavier Chateau.
\newblock Liquid bridge between two moving spheres: an experimental study of
  viscosity effects.
\newblock {\em Journal of colloid and interface science}, 231(1):26--31, 2000.

\bibitem{bozkurt2017capillary}
MG~Bozkurt, Dante Fratta, and WJ~Likos.
\newblock Capillary forces between equally sized moving glass beads: an
  experimental study.
\newblock {\em Canadian Geotechnical Journal}, 54(9):1300--1309, 2017.

\bibitem{likos2004unsaturated}
Ning Lu and William~J Likos.
\newblock Unsaturated soil mechanics.
\newblock {\em ed: John Wiley and Sons Inc., New Jersey}, 2004.

\bibitem{fisher1926capillary}
RA~Fisher.
\newblock On the capillary forces in an ideal soil; correction of formulae
  given by wb haines.
\newblock {\em The Journal of Agricultural Science}, 16(3):492--505, 1926.

\bibitem{derjaguin1934untersuchungen}
Boris Derjaguin.
\newblock Untersuchungen ueber die reibung und adhaesion, iv: Theorie des
  anhaftens kleiner teilchen.
\newblock {\em Kolloid-Zeitschrift}, 69:155--164, 1934.

\bibitem{israelachvili2022surface}
Jacob~N Israelachvili.
\newblock Surface forces.
\newblock In {\em The Handbook of Surface Imaging and Visualization}, pages
  793--816. CRC Press, 2022.

\bibitem{rabinovich2005capillary}
Yakov~I Rabinovich, Madhavan~S Esayanur, and Brij~M Moudgil.
\newblock Capillary forces between two spheres with a fixed volume liquid
  bridge: theory and experiment.
\newblock {\em Langmuir}, 21(24):10992--10997, 2005.

\bibitem{butt2009normal}
Hans-J{\"u}rgen Butt and Michael Kappl.
\newblock Normal capillary forces.
\newblock {\em Advances in colloid and interface science}, 146(1-2):48--60,
  2009.

\bibitem{lian1993theoretical}
Guoping Lian, Colin Thornton, and Michael~J Adams.
\newblock A theoretical study of the liquid bridge forces between two rigid
  spherical bodies.
\newblock {\em Journal of colloid and interface science}, 161(1):138--147,
  1993.

\bibitem{maugis1987adherence}
D~Maugis.
\newblock Adherence of elastomers: Fracture mechanics aspects.
\newblock {\em Journal of Adhesion Science and Technology}, 1(1):105--134,
  1987.

\bibitem{taylor1934formation}
Geoffrey~Ingram Taylor.
\newblock The formation of emulsions in definable fields of flow.
\newblock {\em Proceedings of the Royal Society of London. Series A, containing
  papers of a mathematical and physical character}, 146(858):501--523, 1934.

\bibitem{grace1982dispersion}
Harold~P Grace.
\newblock Dispersion phenomena in high viscosity immiscible fluid systems and
  application of static mixers as dispersion devices in such systems.
\newblock {\em Chemical Engineering Communications}, 14(3-6):225--277, 1982.

\bibitem{bentley1986experimental}
BJ~Bentley and L~Gary Leal.
\newblock An experimental investigation of drop deformation and breakup in
  steady, two-dimensional linear flows.
\newblock {\em Journal of Fluid Mechanics}, 167:241--283, 1986.

\bibitem{stone1989relaxation}
Howard~A Stone and L~Gary Leal.
\newblock Relaxation and breakup of an initially extended drop in an otherwise
  quiescent fluid.
\newblock {\em Journal of Fluid Mechanics}, 198:399--427, 1989.

\bibitem{shenoy2016stokes}
Anish Shenoy, Christopher~V Rao, and Charles~M Schroeder.
\newblock Stokes trap for multiplexed particle manipulation and assembly using
  fluidics.
\newblock {\em Proceedings of the National Academy of Sciences},
  113(15):3976--3981, 2016.

\bibitem{kumar2019orientation}
Dinesh Kumar, Anish Shenoy, Songsong Li, and Charles~M Schroeder.
\newblock Orientation control and nonlinear trajectory tracking of colloidal
  particles using microfluidics.
\newblock {\em Physical Review Fluids}, 4(11):114203, 2019.

\bibitem{shenoy2019flow}
Anish Shenoy, Dinesh Kumar, Sascha Hilgenfeldt, and Charles~M Schroeder.
\newblock Flow topology during multiplexed particle manipulation using a stokes
  trap.
\newblock {\em Physical Review Applied}, 12(5):054010, 2019.

\bibitem{kumar2020automation}
Dinesh Kumar, Anish Shenoy, Jonathan Deutsch, and Charles~M Schroeder.
\newblock Automation and flow control for particle manipulation.
\newblock {\em Current Opinion in Chemical Engineering}, 29:1--8, 2020.

\bibitem{tu20233d}
Michael~Q Tu, Hung~V Nguyen, Elliel Foley, Michael~I Jacobs, and Charles~M
  Schroeder.
\newblock 3d manipulation and dynamics of soft materials in 3d flows.
\newblock {\em Journal of Rheology}, 67(4):877--877, 2023.

\bibitem{lopez2023low}
Ramon Lopez, Jovina Vaswani, Dylan~T Butler, Joseph McCarthy, and Sachin~S
  Velankar.
\newblock Low viscosity liquid bridges: Stretching of liquid bridges immersed
  in a higher viscosity liquid.
\newblock {\em JCIS Open}, 9:100079, 2023.

\bibitem{lian2016capillary}
Guoping Lian and Jonathan Seville.
\newblock The capillary bridge between two spheres: new closed-form equations
  in a two century old problem.
\newblock {\em Advances in colloid and interface science}, 227:53--62, 2016.

\bibitem{graham2018microhydrodynamics}
Michael~D Graham.
\newblock {\em Microhydrodynamics, Brownian Motion, and Complex Fluids}.
\newblock Cambridge University Press, 2018.

\bibitem{vincent1991watersheds}
Luc Vincent and Pierre Soille.
\newblock Watersheds in digital spaces: an efficient algorithm based on
  immersion simulations.
\newblock {\em IEEE Transactions on Pattern Analysis \& Machine Intelligence},
  13(06):583--598, 1991.

\bibitem{preim2013visual}
Bernhard Preim and Charl~P Botha.
\newblock {\em Visual computing for medicine: theory, algorithms, and
  applications}.
\newblock Newnes, 2013.

\bibitem{kimtext}
S.~Kim and S.~J. Karrila.
\newblock {\em Microhydrodynamics: Principles and Selected Applications}.
\newblock Butterworth-Heinemann, 1991.

\bibitem{koura1991variable}
Katsuhisa Koura and Hiroaki Matsumoto.
\newblock Variable soft sphere molecular model for inverse-power-law or
  lennard-jones potential.
\newblock {\em Physics of fluids A: fluid dynamics}, 3(10):2459--2465, 1991.

\bibitem{rabani2001interactions}
Eran Rabani and SA~Egorov.
\newblock Interactions between passivated nanoparticles in solutions: Beyond
  the continuum model.
\newblock {\em The Journal of Chemical Physics}, 115(8):3437--3440, 2001.

\bibitem{liebetreu2020cluster}
Maximilian Liebetreu and Christos~N Likos.
\newblock Cluster prevalence in concentrated ring-chain mixtures under shear.
\newblock {\em Soft Matter}, 16(37):8710--8719, 2020.

\end{thebibliography}

\end{document}